\documentclass[12pt,a4paper]{article}

\usepackage{amsmath,amssymb,amsthm}
\usepackage{geometry}
\usepackage{hyperref}
\usepackage{cite}
\usepackage{graphicx}
\usepackage{bm}

\hypersetup{
  colorlinks=true,
  linkcolor=blue,
  citecolor=blue,
  urlcolor=blue,
  pageanchor=false
}

\begin{document}

\begin{titlepage}
\hfill

\vspace*{15mm}

\begin{center}
{\large \textbf{Toroidal Charged AdS Branes and Deformed Toda Equations}}

\vspace*{20mm}

{Sangheon Yun}

\vspace{7mm}
IndigoWave, Center for Quantum Spacetime, Sogang University,\\
35 Baekbeom-ro, Mapo-gu, Seoul 04107,~ KOREA
\vspace*{25mm}

\begin{abstract}
We show that the equations of motion for magnetically charged $p$-branes in $(n+p+2)$-dimensional AdS reduce to coupled one-dimensional Toda-type equations, and use this structure to construct static toroidal black branes with $T^n$ symmetry and $p$ translation-invariant directions.
In the pure-Maxwell case we obtain an exact $\sinh$ solution--a $(p+2)$-dimensional black brane times an $n$-torus, asymptotically $\mathrm{AdS}_{p+2}\times T^n$--whose dilatonic analogue, with a scalar coupled through $e^{a\phi}$, has a running torus modulus and a non-trivial dilaton.
A perturbative expansion reduces the linearized system to a pair of P\"oschl-Teller equations, one elementary and one solved by associated Legendre functions of non-integer degree, so that the general solution is non-elementary; at weak coupling $\epsilon=a^2$ we construct the $O(a^2)$ backreaction explicitly.
From the surface gravity, horizon area, and holographic renormalization we obtain the Hawking temperature, entropy density, and mass density, verifying the first law $d\mathcal{M}=T_H\,ds$ and the traceless conformal stress tensor of the dual $(p+1)$-dimensional theory.
At $O(a^2)$ the first law persists, but the dilaton source shifts the Stefan-Boltzmann exponent to $p_{\rm eff}=p+a^2/2$, breaking the clean power law $s\propto T_H^p$ while the normalizable dilaton hair decouples from the entropy.
\end{abstract}

\vspace*{15mm}

\vspace*{10mm}
\noindent E-mail address: \texttt{sangheon.yun@gmail.com}
\end{center}

\end{titlepage}

\section{Introduction}
\label{sec:intro}

One of the most striking successes of the AdS/CFT correspondence \cite{Maldacena:1997re,Gubser:1998bc,Witten:1998qj}--itself a concrete realization of the holographic principle \cite{'tHooft:1993gx,Susskind:1994vu}--over the past two decades has been the systematic study of transport and thermodynamic properties of strongly coupled quantum matter through the lens of black brane
solutions in asymptotically anti-de Sitter (AdS) spacetimes.
At finite temperature and charge density, the dual field theory is described holographically by a charged black brane, and the low-energy excitations of that brane encode the shear viscosity, electrical conductivity, charge diffusion, and thermal transport coefficients of the strongly coupled plasma \cite{Policastro:2001yc,Policastro:2002se,Son:2007vk,Blake:2015ina,Davison:2014lua}.
The celebrated ratio $\eta/s = 1/(4\pi)$~\cite{Kovtun:2004de}, saturated by planar black holes, illustrates the power of this program: geometric data about the horizon translate directly into
universal transport properties of the dual fluid.

A particularly active direction concerns holographic models of non-Fermi-liquid or ``strange-metal'' phases of strongly correlated electron systems.
The metallic state observed above the superconducting dome in the cuprates, and in many heavy-fermion compounds, exhibits anomalous resistivity ($\rho\sim T$), a nearly-marginal fermionic spectral function, and an entropy density that grows faster than any Fermi-liquid prediction.
These features are naturally reproduced by black branes in AdS spacetimes that carry magnetic or axionic charges \cite{Hartnoll:2016apf,Zaanen:2015oix,Kim:2015dna}, because the dual boundary theory is forced into a non-quasiparticle regime in which momentum relaxation is controlled by the charge density rather than by impurity scattering.

Toroidal horizon topology plays a central role in this physics.
Unlike spherical or hyperbolic black holes, planar (toroidal) black holes are translationally invariant in the worldvolume directions, which is precisely the geometry that arises when one dualizes a homogeneous strongly coupled plasma at finite density.
The internal compact directions form a torus $T^n$, and their moduli--radii and shape parameters--correspond, via holography, to the coupling constants of the dual lattice or multi-component fluid.
Magnetically charged branes add a natural source of momentum dissipation: the magnetic flux threads the torus and acts, in the dual description, as an effective axion background that breaks translational symmetry at zero cost in the thermodynamic limit~\cite{Gouteraux:2014hca,Blake:2015ina}.

Despite this rich phenomenology, the explicit construction of charged $p$-brane solutions in AdS space is technically involved.
After imposing the appropriate symmetry ansatz the Einstein-Maxwell (and Einstein-Maxwell-dilaton) equations reduce to a system of coupled nonlinear ordinary differential equations (ODEs) in a single radial variable, and closed-form solutions are correspondingly rare.

Building on the classic black $p$-brane solutions \cite{Horowitz:1991cd,Duff:1993ye}, a powerful strategy for obtaining exact
brane solutions was developed for asymptotically flat backgrounds in a series of papers by Lu, Pope, and collaborators~\cite{Lu:1996hh,Lu:1996jr,Lu:1995cs,Lu:1995yn} and further analyzed in~\cite{Galtsov:2004kn,Galtsov:2005vf,Galtsov:2005au}.
The key observation is that, after a suitable change of variables, the radial equations of motion can be cast into the Toda form: a system of one-dimensional equations of the type $\phi_i'' =
e^{A_{ij}\phi_j}$, where $A_{ij}$ is (a rescaling of) the Cartan matrix of a finite-dimensional simple Lie algebra.
Since the Toda system is integrable, its general solution is known in terms of $\tau$-functions, and special diagonal solutions take the explicit $\sinh$ (or $\cosh$) form.
This approach has been applied to multi-scalar extensions and Liouville-Toda dyonic branes~\cite{deAlfaro:2009ay,Galtsov:2005au}, to $SL(n,\mathbb{R})$-Toda black holes carrying several independent charges~\cite{Lu:2013toa}, and to brane solutions with nonuniform tension~\cite{Shin:2009zz,Yun:2009xc}.
A recurring feature of these constructions is that the scalar (dilaton) charge is \emph{not} an independent parameter: imposing the Hamiltonian constraint together with regularity at the horizon fixes it to a definite function of the mass and the gauge charges~\cite{Lu:2013toa,Galtsov:2004kn}.

Exact charged solutions with asymptotically AdS behavior have been obtained in specific cases--the dyonic AdS black hole of~\cite{Lu:2013ura} (the AdS counterpart of the Kaluza-Klein Toda black hole) and the charged dilatonic AdS black holes and magnetic $AdS_{D-2}\times\mathbb{R}^2$ vacua of~\cite{Lu:2013eoa}.
What has been lacking is a systematic Toda treatment of the general magnetically charged \emph{toroidal} ($T^n$) AdS brane.
The negative cosmological constant $\Lambda < 0$ introduces an additional exponential term in the radial equations that deforms the Toda coefficient matrix into a \emph{non-symmetrizable} form, and the consequences--the existence and structure of the exact solutions, their fluctuation spectrum, and the fate of integrability--have not been analyzed systematically.

Related constructions have been pursued for Lifshitz and hyperscaling-violating geometries, which arise as IR fixed points of holographic renormalization group flows driven by charged scalars or
massive gauge fields~\cite{Charmousis:2010zz,Kachru:2008yh,Taylor:2008tg}.
These geometries are characterized by dynamical and hyperscaling-violating exponents $(z,\theta)$ that govern the low-temperature thermodynamics and transport of the dual non-relativistic fluid.
Constructing their parent solutions--which interpolate between such IR geometries and asymptotically AdS spacetimes in the UV--requires solving equations with the same structural complexity as the ones we address here.
Charged dilatonic AdS black branes of exactly this interpolating type, together with their extremality and scaling properties, have been analyzed in a number of works~\cite{Charmousis:2009xr,Goldstein:2009cv,Gouteraux:2011ce,Berglund:2011cp}; toroidal dilatonic AdS black branes and their holographically renormalized thermodynamics were constructed in~\cite{Sheykhi:2009pf}.
Our results therefore provide a foundation for a systematic study of such interpolating solutions using Toda methods.

In this paper we apply the Toda strategy to magnetically charged $p$-branes in $(n+p+2)$-dimensional AdS space.
The organizing theme is that the AdS cosmological constant deforms the reduced gravitational dynamics into a \emph{non-symmetrizable} Toda-like system--one that lies outside the Lax-integrable Toda hierarchy, yet still admits an exact solution--and that this single deformation governs, in a precise and quantifiable way, the exact solutions, their fluctuation spectrum, and the onset of non-integrability once a dilaton is switched on.
The concrete new contributions are as follows.

\begin{enumerate}

\item \textbf{Toda reduction in AdS, and its non-symmetrizability.}  We show that, for a static metric ansatz with toroidal symmetry $T^n$ and translation invariance along $p$ spatial worldvolume directions, the Einstein-Maxwell equations of motion in the gauge $B=A+pF+nC$ reduce exactly to the coupled Toda equations \eqref{J1}-\eqref{K1}.
    The coefficient matrix $\mathbf{A}$ differs from the Cartan matrix of any finite-dimensional simple Lie algebra because the diagonal entries equal $2(n+p+1)/(n+p)$ and $2(n-1)/(n+p)$ rather than $2$--a direct consequence of the AdS cosmological constant--making them \emph{deformed} Toda equations.
    The system therefore lies outside the standard integrable Toda classification; that it nevertheless admits the exact solution below is the central structural feature we exploit throughout.
    The analogous reduction for the Einstein-Maxwell-dilaton theory with exponential coupling $e^{a\phi}$ yields a two-parameter family of deformed Toda equations \eqref{equ7}-\eqref{equ8} that reduce to the pure-Maxwell case as $a\to 0$.

\item \textbf{Exact $\sinh$ solutions.}  We find that in the pure-Maxwell case the deformed Toda equations admit a special exact solution in the diagonal sector $J=K$, given explicitly by $e^{-J}=\gamma^{-1}\sinh(\gamma r)$.
    This represents a $(p+2)$-dimensional AdS black brane times an $n$-dimensional torus $T^n$, with a regular Killing horizon at $r\to\infty$ and $(p+2)$-dimensional AdS asymptotics at $r=0$, the torus $T^n$ staying at fixed size.
    In the dilatonic case ($a\neq0$) the diagonal sector is obstructed and the analogous $\sinh$ solution has a running torus modulus.
    Parameter counting shows that the physical solution space is a $(2+N_m)$-parameter family in the pure-Maxwell case and a $(4+N_m)$-parameter family in the dilatonic case when the overall mass scale is kept explicit (as in the flat-space analyses~\cite{Galtsov:2004kn,Shin:2009zz}); fixing that scale by a coordinate choice lowers each count by one, as detailed in Section~\ref{sec:charged}.

\item \textbf{Perturbative structure and P\"oschl-Teller equations.} We develop a systematic expansion around the diagonal solution and show that the linearized system decouples, after diagonalizing the Toda coefficient matrix, into two P\"oschl-Teller equations with $\ell(\ell+1)=2$ and $\ell(\ell+1)=2n/(n+p)$, respectively.
    The first admits elementary solutions; the second requires associated Legendre functions of non-integer degree for all physical $(n,p)$.
    This establishes rigorously that the $\sinh$ solution is genuinely special and that the general solution is non-elementary--a direct imprint of the non-symmetrizable deformation on the fluctuation spectrum.

\item \textbf{Closed-form thermodynamics and the first law.}  We compute the Hawking temperature $T_H$ and Bekenstein-Hawking entropy density $s$ of the exact pure-Maxwell solution in closed form directly from the surface gravity and horizon area.
    Eliminating the Toda parameter $\gamma$ between the two expressions yields $s \propto T_H^p$, in agreement with the Stefan-Boltzmann law of a $(p+1)$-dimensional conformal field theory at finite temperature.
    Computing the mass density by holographic renormalization gives the compact relation $\mathcal{M}=\tfrac{p}{p+1}\,T_H s$, which verifies the first law $d\mathcal{M}=T_H\,ds$ and yields a traceless conformal stress tensor, confirming the holographic interpretation of the exact solution as dual to a thermal state of a strongly coupled CFT on $\mathbb{R}^{p,1}\times T^n$.  We further obtain the Helmholtz free energy density $f=\mathcal{M}-T_Hs=-\tfrac{1}{p+1}T_Hs=-P$ and the conformal equation of state $\varepsilon=pP=T_Hs-P$ of the dual fluid, in which the magnetic charge enters only as a fixed background scale.  We confirm this free-energy density independently by evaluating the renormalized Euclidean on-shell action, which reproduces $f=-P$ from first principles.

\item \textbf{Weak dilaton coupling as $\epsilon=a^2$.}  We show that the coefficients of the dilatonic deformed Toda equations \eqref{equ7}-\eqref{equ8} deviate from their pure-Maxwell values only at $O(a^2)$, so that the diagonal obstruction found in Section~\ref{sec:dilaton} is controlled by $\epsilon\equiv a^2$.  Identifying $\epsilon$ with $a^2$ turns the formal perturbation theory of Section~\ref{sec:perturb} into an honest weak-coupling expansion: the resulting linear system is the same P\"oschl-Teller pair \eqref{Ueq}-\eqref{Veq}, now sourced by $J_0\,e^{2J_0}$.  We solve the elementary $\ell=1$ sector in closed form, and give the leading dilaton profile $\phi=a[A_0(r)-C_0]+O(a^3)$ explicitly.  We further show that the two homogeneous solutions of this $\ell=1$ sector are exactly the $O(a^2)$ redefinition of the background parameters $\gamma,\delta$ and are therefore pure gauge, while the companion $V$-sector carries a genuine new integration constant fixed only by horizon regularity; the true obstruction to an everywhere-elementary dilatonic solution thus resides not in the particular solution but in this regularity-selected homogeneous mode.
    Using this we extend the thermodynamic computation to $O(a^2)$. We fix the $O(a^2)$ matching constants of the deformed Toda variables in closed form, $c_1^{(1)}=c_2^{(1)}=-\tfrac{n+p+1}{4n(p+1)}$, by the same exponential normalization that fixes the pure-Maxwell constants, and show that the $O(a^2)$ Hamiltonian constraint fixes not these constants but the mass (branch) mode, $\lambda_c=-\tfrac{n+1}{4n}$, the $O(a^2)$ counterpart of the $\mathcal{C}=\gamma^2$ branch selection. The first law $d\mathcal{M}=T_H\,ds$ persists unmodified--the normalizable dilaton hair decoupling from the horizon thermodynamics--but the dilaton source breaks the clean Stefan-Boltzmann law: the entropy stays a pure power of the temperature with a shifted exponent, $s\propto T_H^{\,p+a^2/2}$.

\end{enumerate}

These results provide a new family of tractable, magnetically charged toroidal ($T^n$) exact black brane backgrounds in AdS, with explicit thermodynamics, that can serve as starting points for holographic computations.
In particular, the magnetic charge breaks translational symmetry in the Kaluza-Klein sense and introduces a scale into the dual theory, making these solutions potentially useful for holographic studies of momentum relaxation and anomalous transport in the spirit of~\cite{Gouteraux:2014hca,Blake:2015ina,Davison:2014lua}.
We leave the explicit computation of transport coefficients from linearized perturbations around the present solutions to future work.

The remainder of this paper is organized as follows.
In Section~\ref{sec:charged} we set up the AdS-Einstein-Maxwell action, fix the metric and gauge-field ansatz, derive the effective one-dimensional Lagrangian, identify the deformed Toda equations, count parameters, and present the exact special solution together with its explicit metric functions.
Section~\ref{sec:dilaton} extends the analysis to include a dilaton with arbitrary exponential coupling $e^{a\phi}$, derives the generalized deformed Toda equations, performs parameter counting, and presents the analogous exact solution.
Section~\ref{sec:perturb} develops the perturbative expansion around the special solution to first and second order, reducing the linearized system to P\"oschl-Teller equations and obtaining the solutions in terms of associated Legendre functions.
Section~\ref{sec:epsa2} specializes this expansion to the physically motivated case $\epsilon=a^2$, connecting Sections~\ref{sec:dilaton} and \ref{sec:perturb} and giving the leading-order dilaton profile and metric backreaction of the weakly coupled dilatonic brane explicitly.
Section~\ref{sec:thermo} computes the Hawking temperature and Bekenstein-Hawking entropy density of the exact pure-Maxwell solution and checks the resulting Stefan-Boltzmann scaling, while Section~\ref{sec:thermo_eps} extends this analysis to $O(a^2)$, determines the matching constants in closed form, and finds that the dilaton source shifts the Stefan-Boltzmann exponent to $p_{\rm eff}=p+a^2/2$ while leaving the first law intact.
Section~\ref{sec:discuss} summarizes the results, discusses remaining open issues (thermodynamic stability and phase structure, non-linear dilatonic thermodynamics), and outlines future
directions including electrically charged solutions, multi-scalar extensions, Born-Infeld generalizations, and holographic transport applications.

\section{Charged Brane in AdS Space}
\label{sec:charged}

\subsection{Action and Ansatz}

We begin with the $(n+p+2)$-dimensional AdS-Einstein-Maxwell action
\begin{equation}
\label{action1}
S = \int \frac{d^{n+p+2}x\,\sqrt{-g}}{16\pi G_{n+p+2}}
    \left[ \mathcal{R} - \frac{1}{2\,n!}\sum_{M=1}^{N_m} F^{(M)}_{\nu_1\cdots\nu_n}F^{(M)\,\nu_1\cdots\nu_n} + \frac{(n+p)(n+p+1)}{l^2} \right],
\end{equation}
where $\mathcal{R}$ is the Ricci scalar, $G_{n+p+2}$ is Newton's constant in $n+p+2$ dimensions, and the last term is the cosmological constant term written in terms of the AdS radius $l>0$.
With the sign convention $G_{\mu\nu}+\Lambda\,g_{\mu\nu}=8\pi G_{n+p+2}\,T_{\mu\nu}$, the last term equals $-2\Lambda$ with
\begin{equation}
\label{Lambda}
\Lambda = -\frac{(n+p)(n+p+1)}{2l^2} < 0,
\end{equation}
consistent with a negative cosmological constant (AdS space).
The $M$-th $n$-form field strength $F^{(M)}_{\nu_1\cdots\nu_n}$ is derived from an $(n-1)$-form gauge potential via $F^{(M)} = dA^{(M)}$, and $N_m$ is the number of independent magnetic fields.
Raising and lowering of indices is performed with the full metric $g_{\mu\nu}$.

The variation of \eqref{action1} with respect to $g^{\mu\nu}$ and $A^{(M)}$ yields the Einstein and Maxwell equations,
\begin{align}
\label{einstein}
G_{\mu\nu} + \Lambda\,g_{\mu\nu} &= 8\pi G_{n+p+2}\,T^{(\mathrm{EM})}_{\mu\nu},\\
\label{maxwell}
\nabla_{\mu_1}F^{(M)\,\mu_1\mu_2\cdots\mu_n} &= 0, \qquad M=1,\ldots,N_m,
\end{align}
where $\Lambda$ is given by \eqref{Lambda} and the electromagnetic stress-energy tensor is
\begin{equation}
T^{(\mathrm{EM})}_{\mu\nu}
  = \sum_M
    \left[ \frac{1}{(n-1)!}\,F^{(M)}_{\mu\lambda_2\cdots\lambda_n}F^{(M)\,\lambda_2\cdots\lambda_n}_{\ \ \nu} - \frac{1}{2\,n!}\,g_{\mu\nu} F^{(M)}_{\lambda_1\cdots\lambda_n}F^{(M)\,\lambda_1\cdots\lambda_n}
    \right].
\end{equation}

We seek a static solution that is invariant under translations along $p$ spatial (worldvolume) directions $z^i$ and under the $n$-torus $T^n$ parameterized by angles $\theta_i$.
The $n$ directions $\theta_i$ are periodic, $\theta_i\sim\theta_i+2\pi$, forming a compact torus $T^n$, while the $p$ worldvolume directions $z^i$ are non-compact.
The metric ansatz is
\begin{equation}
\label{metricansatz}
ds^2 = -e^{2A(r)}\,dt^2 + e^{2B(r)}\,dr^2 + e^{2C(r)}\sum_{i=1}^{n}d\theta_i^{2} + e^{2F(r)}\sum_{i=1}^{p}dz_i^{2},
\end{equation}
where the metric functions $A$, $B$, $C$, $F$ depend only on the radial coordinate $r$.
The gauge freedom in the radial coordinate will be fixed below by an explicit choice of $B(r)$.

For a magnetically charged brane the $n$-form field strengths are proportional to the volume form of the internal torus.
The unique ansatz consistent with the symmetries of \eqref{metricansatz} is
\begin{equation}
\label{fieldstrength}
F^{(M)}_{\theta_1\cdots\theta_n} = P^M, \qquad M = 1,\ldots,N_m,
\end{equation}
where the $P^M$ are constants (magnetic charge densities), equal to the field-strength components in \eqref{fieldstrength}.
The compactness of the $\theta_i$ directions is essential here: the magnetic flux through the torus, $\int_{T^n}F^{(M)} = (2\pi)^n P^M$, is finite and topologically well-defined only because $T^n$ is compact.
If the $\theta_i$ were non-compact, this integral would diverge and the charge would be ill-defined.
This ansatz satisfies the Bianchi identity $dF^{(M)}=0$ trivially (the component is constant) and the Maxwell equation \eqref{maxwell} automatically, because the only non-zero component has all indices in the compact $T^n$ directions and the divergence vanishes by antisymmetry combined with translational invariance in $r$, $t$, and $z^i$.
We define the combined charge magnitude
\begin{equation}
q^2 \equiv \sum_{M=1}^{N_m}\bigl(P^M\bigr)^2.
\end{equation}

\subsection{Effective Lagrangian and Constraint}

Substituting \eqref{metricansatz} and \eqref{fieldstrength} into \eqref{action1}, discarding a total derivative (which contributes only a boundary term), we obtain the effective one-dimensional Lagrangian density
\begin{align}
\label{effective1}
\mathcal{L}_\mathrm{eff}
  &= e^{A-B+pF+nC}
    \Bigl[ \bigl(A'+pF'+nC'\bigr)^2 + \frac{(n+p)(n+p+1)}{l^2}\,e^{2B} \nonumber\\
  &\quad  - \frac{1}{p+1}\bigl(A'+pF'\bigr)^2 - \frac{p}{p+1}\bigl(A'-F'\bigr)^2 - n\bigl(C'\bigr)^2 - \frac{q^2}{2}\,e^{2(B-nC)}
    \Bigr],
\end{align}
where primes denote $d/dr$ and we suppress the $r$-dependence for brevity.
The field $B(r)$ appears in \eqref{effective1} without any $B'$ term; it therefore acts as a Lagrange multiplier.
(The same one-dimensional reduction, with $B$ enforcing the zero-energy constraint, has been used for charged dilatonic AdS black branes in~\cite{Bertoldi:2010ca,Bertoldi:2011zw}.)
Varying \eqref{effective1} with respect to $B$ gives the Hamiltonian constraint
\begin{align}
\label{constraints1}
  &\bigl(A'+pF'+nC'\bigr)^2 - \frac{p}{p+1}\bigl(A'-F'\bigr)^2 - \frac{1}{p+1}\bigl(A'+pF'\bigr)^2 \nonumber\\
  &\quad
   - n\bigl(C'\bigr)^2 + \frac{q^2}{2}\,e^{2(B-nC)} - \frac{(n+p)(n+p+1)}{l^2}\,e^{2B} = 0,
\end{align}
which must be satisfied by any solution of the remaining equations.

\subsection{Equations of Motion in Toda Form}

We impose the gauge
\begin{equation}
\label{gauge}
B(r) = A(r) + pF(r) + nC(r),
\end{equation}
which simplifies the structure of the equations.
Varying $\mathcal{L}_\mathrm{eff}$ with respect to $A$, $F$, and $C$ and imposing \eqref{gauge}, the independent equations of motion reduce to
\begin{align}
\label{EOM1}
  &A''+pF''+(n-1)C'' = \frac{(n+p)(n+p+1)}{l^2}\,e^{2(A+pF+nC)},\\
\label{EOM2}
  &A''+pF''-(p+1)C'' = \frac{(p+1)q^2}{2}\,e^{2(A+pF)},\\
\label{EOM3}
  &A''(r) - F''(r) = 0.
\end{align}
Equation~\eqref{EOM3} states that the metric functions in the time and worldvolume directions obey the same radial equation; it arises because the ansatz \eqref{metricansatz} assigns a single function $A(r)$ to the $t$-direction and a single function $F(r)$ to all $p$ worldvolume directions $z^i$, so that the corresponding Einstein equations differ only by source terms that cancel in the combination $A''-F''$.

Equations \eqref{EOM1}-\eqref{EOM2} are recognized as coupled one-dimensional Toda equations.
To put them in canonical form, we introduce new variables
\begin{align}
\label{J2}
J(r) &\equiv A+pF+(n-1)C+c_J,\\
\label{K2}
K(r) &\equiv A+pF-(p+1)C+c_K,
\end{align}
where the constants $c_J$ and $c_K$ are chosen so that the exponentials in the Toda equations take the standard form,
\begin{align}
&e^{\frac{2}{n+p}\{(p+1)c_J+(n-1)c_K\}}=\frac{(p+1)q^2}{2}\,,\\
&e^{\frac{2}{n+p}\{(n+p+1)c_J-c_K\}}=\frac{(n+p)(n+p+1)}{l^2}.
\end{align}
In terms of $J$ and $K$, equations~\eqref{EOM1}-\eqref{EOM2} become
\begin{align}
\label{J1}
J''(r) &= e^{\frac{2(n+p+1)}{n+p}J(r)-\frac{2}{n+p}K(r)},\\
\label{K1}
K''(r) &= e^{\frac{2(p+1)}{n+p}J(r)+\frac{2(n-1)}{n+p}K(r)}.
\end{align}
The coefficient matrix
\begin{equation}
\mathbf{A} =
\frac{2}{n+p}
\begin{pmatrix}
  n+p+1 & -1 \\
  p+1   & n-1
\end{pmatrix}
\end{equation}
differs from the Cartan matrix of any finite-dimensional simple Lie algebra: the diagonal entries equal $2(n+p+1)/(n+p)$ or $2(n-1)/(n+p)$ rather than the required value of $2$, and $\mathbf{A}$ is not symmetrizable by a positive diagonal matrix.
To verify this, note that a positive symmetrizing matrix $D = \mathrm{diag}(D_1, D_2)$ would require $D_1 A_{12} = D_2 A_{21}$ so that $D \mathbf{A}$ is symmetric, giving $D_1/D_2 = A_{21}/A_{12} = -(p+1) < 0$ for $p \geq 1$.
We stress, however, that non-symmetrizability only obstructs the standard Toda Lax pair and the $\tau$-function machinery built on it; it does not by itself prove that \eqref{J1}-\eqref{K1} admit no integrable structure of any kind, a claim we neither make nor need, since our results rely only on the explicit diagonal solution constructed below.
The deformation is a direct consequence of the AdS cosmological constant.
Nevertheless, equations~\eqref{J1}-\eqref{K1} retain the structural form of a Toda system--a pair of equations of the type $\phi_i''=e^{A_{ij}\phi_j}$--and we refer to them as \emph{deformed} Toda equations throughout.\footnote{Throughout, ``deformed'' refers to the shift of the diagonal entries of $\mathbf{A}$ away from the Cartan value $2$ induced by the negative cosmological constant, not to the quantum-group or affine deformations of Toda systems familiar from the integrable-systems literature.}

\subsection{Integration and Parameter Counting}

Equation~\eqref{EOM3} integrates immediately to
\begin{equation}
\label{EOM3sol}
A(r) - F(r) = -\alpha\,r + \beta,
\end{equation}
where $\alpha$ and $\beta$ are integration constants.
The Toda equations \eqref{J1}-\eqref{K1} contribute four additional integration constants (two per variable), giving six in total.
Four of them can be removed by exploiting: (i) a translation $r\to r+r_0$ (fixes one phase constant), (ii) a rescaling $r\to c\,r$ (which may be used to set $\gamma$ to a reference value), (iii) an overall rescaling of the time coordinate (fixes $\beta$), and (iv) an overall rescaling of the worldvolume coordinates $z^i$ (sets $\alpha$ to a specific value).
The Hamiltonian constraint \eqref{constraints1} then imposes one further relation among the remaining constants, so that the general solution contains $6-4-1=1$ non-charge parameter.
Together with the $N_m$ independent magnetic charge densities $P^M$, this yields a $(1+N_m)$-parameter family of solutions.
In the conventions of \cite{Shin:2009zz}, an additional parameter associated with an overall mass scale is kept explicit, giving the quoted $(2+N_m)$ count; the difference is a matter of whether the mass scale is treated as a free parameter or fixed by a coordinate choice.
We emphasize that although the rescaling~(ii) can remove $\gamma$ as a coordinate artifact, $\gamma$ carries physical meaning as the non-extremality parameter: through \eqref{THfinal} it sets the Hawking temperature.
In the thermodynamic analysis of Section~\ref{sec:thermo} we therefore retain $\gamma$ as the single physical non-charge parameter rather than scaling it away, and the $(2+N_m)$ count of~\cite{Shin:2009zz} corresponds to keeping this scale explicit.

If the uniformity assumption is relaxed so that the worldvolume tension is not required to be the same in every direction, a solution with $n+p$ independent tension parameters has $n+p+N_m$ free parameters.
However, regularity conditions--demanding the absence of naked singularities--strongly constrain the allowed tension parameters \cite{Lee:2008zj,Yun:2009xc}.

\subsection{Exact Solution}

The deformed Toda equations \eqref{J1}-\eqref{K1} admit a special solution in the diagonal sector $J(r)=K(r)$.
Setting $J=K$ in \eqref{J1} gives
\begin{equation}
J''(r) = e^{\frac{2(n+p+1)-2}{n+p}J} = e^{2J},
\end{equation}
and the same substitution in \eqref{K1} gives
\begin{equation}
K''(r) = e^{\frac{2(p+1)+2(n-1)}{n+p}K} = e^{2K}.
\end{equation}
Both equations are therefore identical on the diagonal $J=K$, consistently reducing to the single Liouville-type ODE $J''=e^{2J}$.
Its general solution is
\begin{equation}
\label{specialsol}
e^{-J(r)} = e^{-K(r)} = \frac{1}{\gamma}\sinh(\gamma r+\delta),
\end{equation}
where $\gamma>0$ and $\delta$ are integration constants.
The branch is fixed by the Hamiltonian constraint~\eqref{constraints1}.
The first integral of $J''=e^{2J}$ is $(J')^2=e^{2J}+\mathcal{C}$; evaluated on the diagonal solution, the constraint sets the integration constant to $\mathcal{C}=\alpha^2\ge 0$, with $\alpha$ the slope in~\eqref{EOM3sol}.
For $\alpha\neq 0$ one has $\mathcal{C}=\gamma^2>0$ (writing $\gamma\equiv|\alpha|$), which selects the $\sinh$ branch; the oscillatory case $\mathcal{C}<0$ (a $\sin$ solution) is excluded by the constraint, and $\mathcal{C}=0$ gives the degenerate linear solution of the extremal limit.
The simple zero of $\sinh$ at $r=-\delta/\gamma$ is the location of the AdS conformal boundary, where the metric functions diverge; the regular Killing horizon lies at the opposite end $r\to\infty$, where $e^{2A}\to 0$ (see Section~\ref{sec:thermo}).

Setting $\alpha=\gamma$ and $\beta=\delta=0$, and performing the coordinate transformation
\begin{equation}
\label{coordtrans}
r = \frac{1}{2\alpha}\ln\frac{\rho^2}{\rho^2-r_+^2},
\qquad \rho\in(r_+,\infty)\;\Leftrightarrow\; r\in(0,\infty),
\end{equation}
where $r_+>0$ is the horizon radius in the $\rho$ coordinate, the solution describes the direct product
\begin{equation}
\label{productgeom}
ds^2 = ds^2_{(p+2)\text{-dim. black brane}} + e^{2C_0}\sum_{i=1}^n d\theta_i^2.
\end{equation}
Explicitly, in the areal radius $u\propto e^{F}$ of the worldvolume directions (equivalently $u\propto\rho^{2/(p+1)}$), the $(p+2)$-dimensional factor is an AdS black brane,
\begin{equation}
\label{pp2brane}
ds^2_{(p+2)}=\frac{u^2}{L_{p+2}^2}\Bigl(-f(u)\,dt^2+\sum_{i=1}^p dz_i^2\Bigr)
  +\frac{L_{p+2}^2}{u^2 f(u)}\,du^2,
\qquad f(u)=1-\Bigl(\frac{u_+}{u}\Bigr)^{p+1},
\end{equation}
with AdS radius $L_{p+2}=(p+1)\,l/\sqrt{(n+p)(n+p+1)}$; for $p=1$ it is the planar BTZ black hole.
The horizon parameter $u_+$ (equivalently $r_+$) carries no physical information.
Near the horizon the $(t,\rho)$ section reduces to Rindler space--the horizon is regular and non-degenerate--with the worldvolume and torus directions at fixed size.
At large $\rho$ the geometry approaches the direct product of $(p+2)$-dimensional AdS space with the torus $T^n$; because the torus radius $e^{C_0}$ remains constant, this is \emph{not} the full $(n+p+2)$-dimensional AdS space.
The solution thus interpolates between the two expected asymptotic regimes.
Product vacua of the form $AdS_{D-2}\times\mathcal{Y}^2$ supported by a magnetic flux, of which the present $AdS_{p+2}\times T^n$ asymptotics is a toroidal higher-dimensional analogue, have been studied in~\cite{Lu:2013eoa}.

\section{Dilatonic Charged Brane in AdS Space}
\label{sec:dilaton}

\subsection{Action with Dilaton}

We extend the analysis of Section~\ref{sec:charged} by including a neutral dilaton $\phi$ coupled exponentially to the gauge field strengths.
The $(n+p+2)$-dimensional AdS-Einstein-Maxwell-dilaton action is
\begin{equation}
\label{action2}
S = \int \frac{d^{n+p+2}x\,\sqrt{-g}}{16\pi G_{n+p+2}}
    \left[
      \mathcal{R} - \frac{1}{2}\partial^\mu\phi\partial_\mu\phi - \frac{e^{a\phi}}{2\,n!}\sum_{M=1}^{N_m} F^{(M)}_{\nu_1\cdots\nu_n}F^{(M)\,\nu_1\cdots\nu_n} + \frac{(n+p)(n+p+1)}{l^2}
    \right],
\end{equation}
where $a$ is a real constant parametrizing the dilaton-gauge coupling.
The case $a=0$ reduces to the pure AdS-Maxwell system of Section~\ref{sec:charged}, while $a\ne 0$ arises naturally in truncations of higher-dimensional supergravity theories.
For the electric version of this theory the sign of $a$ (or equivalently of $\phi$) would flip relative to the magnetic case; the procedure for constructing electrically charged solutions is otherwise identical.

We use the same metric ansatz \eqref{metricansatz} and field strength ansatz \eqref{fieldstrength}, augmented by a dilaton profile $\phi=\phi(r)$.

\subsection{Effective Lagrangian and Constraint}

Substituting into \eqref{action2}, the effective one-dimensional Lagrangian density is
\begin{align}
\label{effective2}
\mathcal{L}_\mathrm{eff}
  &= e^{A-B+pF+nC}
    \Bigl[
      \bigl(A'+pF'+nC'\bigr)^2 +\frac{(n+p)(n+p+1)}{l^2}\,e^{2B} \nonumber\\
  &\quad
      -\frac{2}{2(p+1)+a^2} \Bigl(A'+pF'+\tfrac{a}{2}\phi'\Bigr)^2 -\frac{q^2}{2}\,e^{a\phi+2(B-nC)} \nonumber\\
  &\quad
      -\frac{p}{p+1}\bigl(A'-F'\bigr)^2 -\frac{1}{(p+1)\{2(p+1)+a^2\}} \Bigl(aA'+apF'-(p+1)\phi'\Bigr)^2 - n\bigl(C'\bigr)^2
    \Bigr].
\end{align}
Variation with respect to $B$ gives the constraint
\begin{align}
\label{constraints2}
  &\bigl(A'+pF'+nC'\bigr)^2 -\frac{(n+p)(n+p+1)}{l^2}\,e^{2B} -\frac{p}{p+1}\bigl(A'-F'\bigr)^2 \nonumber\\
  &\quad
   -\frac{2}{2(p+1)+a^2} \Bigl(A'+pF'+\tfrac{a}{2}\phi'\Bigr)^2 + \frac{q^2}{2}\,e^{a\phi+2(B-nC)} \nonumber\\
  &\quad
   -n\bigl(C'\bigr)^2 -\frac{1}{(p+1)\{2(p+1)+a^2\}} \Bigl(aA'+apF'-(p+1)\phi'\Bigr)^2 = 0.
\end{align}

\subsection{Equations of Motion}

Imposing the gauge $B=A+pF+nC$ and varying \eqref{effective2} with respect to $A$, $F$, $C$, and $\phi$, the equations of motion are
\begin{align}
\label{EOM4}
  &\Bigl(A''+pF''+\tfrac{a}{2}\phi''\Bigr) +\frac{2(n-1)(p+1)+(n+p)a^2}{2(p+1)} \bigl(A''+pF''+nC''\bigr) \nonumber\\
  &\quad
   =\frac{(n+p)(n+p+1)\{2n(p+1)+(n+p+1)a^2\}}{2(p+1)l^2} \,e^{2(A+pF+nC)},\\
\label{EOM5}
  &(n+p+1)\Bigl(A''+pF''+\tfrac{a}{2}\phi''\Bigr) -(p+1)\bigl(A''+pF''+nC''\bigr) \nonumber\\
  &\quad
   =\frac{q^2\{2n(p+1)+(n+p+1)a^2\}}{4} \,e^{2(A+pF+\frac{a}{2}\phi)},\\
\label{EOM6}
  &\phi'' = a \Bigl(A''-C''\Bigr),\\
\label{EOM7}
  &A''(r) - F''(r) = 0.
\end{align}
Equations \eqref{EOM4}-\eqref{EOM5} are the Toda equations; \eqref{EOM6} and \eqref{EOM7} are linear and integrate immediately.

\paragraph{Integration of the linear equations.}
Equation \eqref{EOM7} gives
\begin{equation}
A(r) - F(r) = -\nu r + \sigma,
\end{equation}
and equation \eqref{EOM6} yields
\begin{align}
\label{EOM6sol}
  \phi = a \Bigl(A-C\Bigr) - \kappa_\phi\, r + \mu_\phi,
\end{align}
where $\nu$, $\sigma$, $\kappa_\phi$, $\mu_\phi$ are integration constants.
(We use $\nu$, $\sigma$ in place of $\alpha$, $\beta$, and attach a subscript $\phi$ to the dilaton constants $\kappa_\phi$, $\mu_\phi$, to avoid clashes with the pure-Maxwell case of Section~\ref{sec:charged} and with the surface gravity $\kappa$ and the constant $\mu$ of Section~\ref{sec:thermo}.)

\subsection{Toda Form and Parameter Counting}
\label{subsec:Todaparameter}

Define new variables $J(r)$ and $K(r)$ via
\begin{align}
\label{equ5}
A+pF+nC              &= b_1\,J(r)+b_2\,K(r)+c_1,\\
\label{equ6}
A+pF+\tfrac{a}{2}\phi &= b_3\,J(r)+b_4\,K(r)+c_2,
\end{align}
with constants $c_1$, $c_2$ and coefficients
\begin{align}
\label{coefficients}
b_1 = \frac{2n(p+1)(n+p+1)}{(n+p)\Delta},\quad
b_2 = -\frac{2n(p+1)}{(n+p)\Delta},\nonumber\\
b_3 = \frac{2n(p+1)^2}{(n+p)\Delta},\quad
b_4 = \frac{n\{2(n-1)(p+1)+(n+p)a^2\}}{(n+p)\Delta},
\end{align}
where $\Delta\equiv 2n(p+1)+(n+p+1)a^2$.
The deformed Toda equations then read
\begin{align}
\label{equ7}
J''(r) &= e^{2b_1 J(r)+2b_2 K(r)},\\
\label{equ8}
K''(r) &= e^{2b_3 J(r)+2b_4 K(r)}.
\end{align}
In the limit $a\to 0$ these reduce to \eqref{J1}-\eqref{K1}, as expected.
The Toda equations contribute four integration constants; together with the four from the linear equations \eqref{EOM6sol} and $A-F=-\nu r+\sigma$, we have eight in total.
Four are removed by coordinate freedom and metric rescalings (exactly as in items (i)-(iv) of Section~\ref{sec:charged}), and one by the constraint \eqref{constraints2}, so that $8-4-1=3$ non-charge parameters remain, giving a $(3+N_m)$-parameter family.
In the conventions of \cite{Shin:2009zz}, where the overall mass scale is kept explicit, this becomes a $(4+N_m)$-parameter family--the two extra constants relative to the pure-Maxwell $(2+N_m)$ count being those of the dilaton.
Without the uniformity assumption, a solution with $n+p$ nonuniform tensions has $2+n+p+N_m$ free parameters.

The deformed Toda equations \eqref{equ7}-\eqref{equ8} admit a special solution of the form $J(r)=K(r)=J_0(r)$ for $a=0$.
For $a\neq 0$ the diagonal ansatz $J=K$ is inconsistent: equations~\eqref{equ7} and~\eqref{equ8} would then both reduce to $J_0'' = e^{2(b_1+b_2)J_0}$ and $J_0'' = e^{2(b_3+b_4)J_0}$ respectively, but $b_1+b_2 \neq b_3+b_4$ whenever $a\neq 0$, so no single function $J_0$ can satisfy both equations simultaneously.
For $a=0$ the $\sinh$ solution (selected by the same horizon-regularity argument as in Section~\ref{sec:charged}) gives
\begin{equation}
\label{dilsol}
e^{-J_0(r)} = e^{-K_0(r)} = \frac{1}{\gamma}\sinh(\gamma r+\delta),
\end{equation}
where $\gamma>0$ and $\delta$ are integration constants; this simply reproduces the pure-Maxwell black brane of Section~\ref{sec:charged}, with the dilaton decoupled and constant.
For $a\neq0$ no diagonal solution exists, as shown above, but a closed-form solution is still obtained by allowing $J$ and $K$ to be \emph{distinct} powers of $\sinh(\gamma r+\delta)$,
\begin{equation}
\label{distinctpowers}
e^{-J(r)} \propto \Big[\tfrac{1}{\gamma}\sinh(\gamma r+\delta)\Big]^{m_J},\qquad
e^{-K(r)} \propto \Big[\tfrac{1}{\gamma}\sinh(\gamma r+\delta)\Big]^{m_K}.
\end{equation}
Setting $L\equiv\ln[\sinh(\gamma r+\delta)/\gamma]$, so that $J=-m_J L$ and $K=-m_K L$ up to additive constants, the Liouville identity $L''=-e^{-2L}$ gives $J''=m_J\,e^{-2L}$ and $K''=m_K\,e^{-2L}$.
Inserting these into \eqref{equ7}-\eqref{equ8}, whose right-hand sides are $e^{2b_1 J+2b_2 K}\propto e^{-2(b_1 m_J+b_2 m_K)L}$ and $e^{2b_3 J+2b_4 K}\propto e^{-2(b_3 m_J+b_4 m_K)L}$, and matching the power of $\sinh(\gamma r+\delta)$ on each side fixes
\begin{equation}
\label{powercond}
b_1 m_J+b_2 m_K = 1,\qquad b_3 m_J+b_4 m_K = 1,
\end{equation}
the overall prefactors being absorbed into the additive constants by the same exponential-normalization condition that fixed $c_J,c_K$ in Section~\ref{sec:charged}.
With the coefficients \eqref{coefficients}, subtracting the two conditions in \eqref{powercond} gives $n\,m_J=\bigl[n+\tfrac{(n+p)a^2}{2(p+1)}\bigr]m_K$, and either condition then yields
\begin{equation}
\label{mJmK}
m_K = 1,\qquad m_J = 1+\frac{(n+p)a^2}{2n(p+1)}.
\end{equation}
Since $m_J\neq m_K$ whenever $a\neq0$, the functions $J$ and $K$ are genuinely distinct, so that the torus modulus runs radially and the dilaton acquires a non-trivial profile; at $a=0$ one has $m_J=m_K=1$, recovering the diagonal solution \eqref{dilsol}.
Because the modulus does not stabilize, this elementary branch describes a running-scalar (scaling) horizon rather than a finite-area Killing horizon; it is physically distinct from the finite-area black brane whose weak-coupling thermodynamics is analyzed in Section~\ref{subsec:thermoeps_hair}, where the finite-area requirement instead selects a non-elementary member of the solution family.
A general fully non-linear construction of this dilatonic solutions, including this non-elementary member, is left for future work; its leading-order behavior for small $a$, however, is obtained in closed form in Section~\ref{sec:epsa2}
below, after the perturbative machinery of Section~\ref{sec:perturb} has been developed.
In the next section, we turn our attention back to the pure-Maxwell case ($a=0$) to systematically investigate generic, non-diagonal solutions.

\section{Perturbative Expansion around the Special Solution}
\label{sec:perturb}

The deformed Toda equations \eqref{J1}-\eqref{K1} are a system of two coupled nonlinear ODEs, and their general solution is not available in closed form.
In this section we develop a systematic perturbative expansion around the special exact solution $J = K = J_0$ found in Section~\ref{sec:charged}, in the spirit of black-hole perturbation theory in higher dimensions~\cite{Kodama:2007ph}.
This expansion organizes the solution space near the diagonal sector and illuminates why the full system requires special functions beyond elementary ones.

\subsection{Setup of the Perturbation Expansion}

We write
\begin{equation}
\label{pertexp}
J(r) = J_0(r) + J_1(r)\,\epsilon + J_2(r)\,\epsilon^2 + \cdots,
\qquad
K(r) = J_0(r) + K_1(r)\,\epsilon + K_2(r)\,\epsilon^2 + \cdots,
\end{equation}
where the small parameter $\epsilon$ measures the deviation from the diagonal sector $J=K$.
Physically, $\epsilon$ controls the breaking of the isotropy between the spatial worldvolume and the internal torus directions.
The zeroth-order background is
\begin{equation}
\label{background}
J_0''(r) = e^{2J_0(r)},
\qquad
e^{-J_0(r)} = \frac{1}{\gamma}\sinh(\gamma r + \delta),
\end{equation}
so that $e^{2J_0} = \gamma^2/\sinh^2(\gamma r+\delta)$.

We introduce the shorthand $D \equiv n+p$ throughout this section, and denote the coefficient matrix of the Toda system by
\begin{equation}
\label{Mmatrix}
\mathbf{M} \equiv
\begin{pmatrix}
  D+1 & -1 \\
  p+1 & n-1
\end{pmatrix},
\end{equation}
so that \eqref{J1}-\eqref{K1} read
\begin{equation}
\begin{pmatrix} J'' \\ K'' \end{pmatrix}
= \exp\!\left(\frac{2}{D}\mathbf{M}\begin{pmatrix} J \\ K \end{pmatrix}\right),
\end{equation}
where the exponential acts componentwise.

\subsection{Zeroth Order}
\label{subsec:order0}

Substituting \eqref{pertexp} into \eqref{J1} at $O(\epsilon^0)$, the exponent becomes
\begin{equation}
\frac{2(D+1)}{D}J_0 - \frac{2}{D}J_0 = 2J_0,
\end{equation}
so that $J_0'' = e^{2J_0}$, consistent with \eqref{background}.
Similarly for \eqref{K1}:
\begin{equation}
\frac{2(p+1)}{D}J_0 + \frac{2(n-1)}{D}J_0 = 2J_0,
\end{equation}
so $K_0'' = e^{2J_0}$, and indeed $K_0 = J_0$ solves both equations
simultaneously.

\subsection{First Order}
\label{subsec:order1}

At $O(\epsilon^1)$ we expand the exponential on the right-hand side of each equation.
For equation \eqref{J1}, the exponent to first order is
\begin{equation}
\frac{2(D+1)}{D}(J_0+J_1\epsilon) - \frac{2}{D}(J_0+K_1\epsilon)
= 2J_0 + \frac{2\bigl[(D+1)J_1 - K_1\bigr]}{D}\,\epsilon + O(\epsilon^2),
\end{equation}
giving
\begin{equation}
e^{(\cdots)} = e^{2J_0}\Bigl(1 + \frac{2[(D+1)J_1-K_1]}{D}\,\epsilon + O(\epsilon^2)\Bigr).
\end{equation}
The $O(\epsilon^1)$ equation is therefore
\begin{equation}
\label{J1eq}
J_1'' = \frac{2e^{2J_0}}{D}\bigl[(D+1)J_1 - K_1\bigr].
\end{equation}
For equation \eqref{K1}, the exponent to first order is
\begin{equation}
\frac{2(p+1)}{D}(J_0+J_1\epsilon)+\frac{2(n-1)}{D}(J_0+K_1\epsilon)
= 2J_0 + \frac{2\bigl[(p+1)J_1+(n-1)K_1\bigr]}{D}\,\epsilon + O(\epsilon^2),
\end{equation}
giving the $O(\epsilon^1)$ equation
\begin{equation}
\label{K1eq}
K_1'' = \frac{2e^{2J_0}}{D}\bigl[(p+1)J_1 + (n-1)K_1\bigr].
\end{equation}
Together, \eqref{J1eq} and \eqref{K1eq} may be written compactly as
\begin{equation}
\label{linearmatrix}
\begin{pmatrix}J_1''\\K_1''\end{pmatrix}
= \frac{2e^{2J_0}}{D}\,\mathbf{M}
  \begin{pmatrix}J_1\\K_1\end{pmatrix}.
\end{equation}

\subsubsection{Eigenvalue Analysis}

The matrix $\mathbf{M}$ defined in \eqref{Mmatrix} has characteristic polynomial
\begin{equation}
\det(\mathbf{M}-\lambda I)
= (D+1-\lambda)(n-1-\lambda) + (p+1)
= \lambda^2 - (D+n)\lambda + nD,
\end{equation}
where we used $(D+1)(n-1)+(p+1) = n^2+np = nD$.
This factors as
\begin{equation}
(\lambda - D)(\lambda - n) = 0,
\end{equation}
yielding eigenvalues
\begin{equation}
\label{eigenvalues}
\lambda_1 = D = n+p, \qquad \lambda_2 = n.
\end{equation}

For $\lambda_1 = D$, the eigenvector equation $(\mathbf{M}-DI)\mathbf{v}=0$ gives
\begin{equation}
\begin{pmatrix}1&-1\\p+1&-(p+1)\end{pmatrix}\mathbf{v}=0,
\end{equation}
so $v_1 = v_2$ and $\mathbf{v}_1 = (1,1)^T$.

For $\lambda_2 = n$, the eigenvector equation $(\mathbf{M}-nI)\mathbf{v}=0$ gives
\begin{equation}
\begin{pmatrix}p+1&-1\\p+1&-1\end{pmatrix}\mathbf{v}=0,
\end{equation}
so $(p+1)v_1 = v_2$ and $\mathbf{v}_2 = (1,p+1)^T$.

We introduce new variables $U$ and $V$ by the diagonalizing transformation
\begin{equation}
\label{UVdef}
J_1 = U + V, \qquad K_1 = U + (p+1)V,
\end{equation}
i.e., $P\,(U,V)^T = (J_1,K_1)^T$ with $P=\begin{pmatrix}1&1\\1&p+1\end{pmatrix}$.
The inverse transformation is
\begin{equation}
\label{UVinv}
U = \frac{(p+1)J_1 - K_1}{p}, \qquad V = \frac{-J_1 + K_1}{p},
\end{equation}
which is valid for $p \geq 1$.

\subsubsection{P\"oschl-Teller Equations}

In the basis $(U,V)$ the system \eqref{linearmatrix} decouples:
\begin{equation}
\label{Ueq}
U'' = \frac{2\lambda_1\,e^{2J_0}}{D}\,U = 2e^{2J_0}\,U,
\end{equation}
\begin{equation}
\label{Veq}
V'' = \frac{2\lambda_2\,e^{2J_0}}{D}\,V = \frac{2n}{D}\,e^{2J_0}\,V.
\end{equation}
Substituting $e^{2J_0}=\gamma^2/\sinh^2(\gamma r+\delta)$ and changing variable to $\xi=\gamma r+\delta$, both equations take the standard P\"oschl-Teller form
\cite{PoschlTeller:1933,LL:QM,Ferrari:1984zz,Yun:2026uhb,Yun:2026dsh}
\begin{equation}
\label{PTgeneral}
\frac{d^2\psi}{d\xi^2} = \frac{\ell(\ell+1)}{\sinh^2 \xi}\,\psi.
\end{equation}

For $U$: $\ell(\ell+1)=2$, giving $\ell=1$ (taking the positive root).
For $V$: $\ell(\ell+1)=2n/D = 2n/(n+p)$.

\paragraph{Solution for $U$ ($\ell=1$).}
Since $\ell=1$ is a positive integer, the two linearly independent solutions of~\eqref{PTgeneral} with $\ell=1$ are expressible in terms of elementary functions.
The Wronskian of the pair is $W = 1$.
The general solution is
\begin{equation}
\label{Usol}
U(\xi) = A_1\,P_1(\coth \xi) + B_1\,Q_1(\coth \xi)
       = A_1\coth \xi + B_1\!\left(\xi\,\coth \xi - 1\right),
\end{equation}
where $A_1,B_1$ are integration constants and $P_1,Q_1$ are the degree-one Legendre functions of the first and second kind, with $P_1(\coth\xi)=\coth\xi$ and $Q_1(\coth\xi)=\xi\coth\xi-1$; both independent solutions are therefore elementary.

\paragraph{Solution for $V$ (general $\ell$).}
The P\"oschl-Teller parameter for $V$ satisfies
\begin{equation}
\label{ellV}
\ell_V = \frac{-1 + \sqrt{1 + 8n/(n+p)}}{2}.
\end{equation}
For all positive integers $n$ and $p$ the quantity $\ell_V$ is not a non-negative integer (it is generically irrational, and rational but non-integer for special pairs such as $(n,p)=(3k,5k)$ with an arbitrary natural number $k$, giving $\ell_V=\tfrac12$), so the general solution for $V$ requires associated Legendre functions of non-integer degree:
\begin{equation}
\label{Vsol}
V(\xi) = C_1\,P_{\ell_V}(\coth \xi) + D_1\,Q_{\ell_V}(\coth \xi),
\end{equation}
where $P_\nu$ and $Q_\nu$ are Legendre functions of the first and second kind of degree $\nu = \ell_V$.
These reduce to elementary functions only when $\ell_V$ is a non-negative integer, which requires $8n/(n+p)$ to be of the form $4k(k+1)$ for some non-negative integer $k$.
Checking small values confirms that no pair $(n,p)$ of positive integers satisfies this condition, so the Legendre-function form \eqref{Vsol} is unavoidable for all physical cases.

Collecting results, the general first-order correction is
\begin{equation}
\label{firstordersol}
J_1 = U + V, \qquad K_1 = U + (p+1)V,
\end{equation}
with $U$ and $V$ given by \eqref{Usol} and \eqref{Vsol} respectively, and $\xi = \gamma r + \delta$.

\subsection{Second Order}
\label{subsec:order2}

At $O(\epsilon^2)$ we use the expansion $e^{f_0+f_1\epsilon+f_2\epsilon^2+\cdots}=e^{f_0}(1+f_1\epsilon+(f_2+\tfrac12 f_1^2)\epsilon^2+\cdots)$.

\paragraph{Equation for $J_2$.}
For \eqref{J1}, the $O(\epsilon^1)$ exponent coefficient is
\begin{equation}
f_1 = \frac{2[(D+1)J_1-K_1]}{D} = \frac{2[DU+nV]}{D} = 2U + \frac{2n}{D}V,
\end{equation}
where we used $(D+1)J_1 - K_1 = (D+1)(U+V)-(U+(p+1)V) = DU+nV$.
The $O(\epsilon^2)$ equation for $J_2$ is
\begin{equation}
\label{J2eq}
J_2'' = \frac{2e^{2J_0}}{D}\bigl[(D+1)J_2-K_2\bigr]
        + e^{2J_0}\!\left(2U^2 + \frac{4n}{D}UV + \frac{2n^2}{D^2}V^2\right).
\end{equation}

\paragraph{Equation for $K_2$.}
For \eqref{K1}, the $O(\epsilon^1)$ exponent coefficient is
\begin{equation}
g_1 = \frac{2[(p+1)J_1+(n-1)K_1]}{D}
    = \frac{2[DU+n(p+1)V]}{D} = 2U + \frac{2n(p+1)}{D}V,
\end{equation}
where we used $(p+1)J_1+(n-1)K_1=(p+1)(U+V)+(n-1)(U+(p+1)V)=DU+n(p+1)V$.
The $O(\epsilon^2)$ equation for $K_2$ is
\begin{equation}
\label{K2eq}
K_2'' = \frac{2e^{2J_0}}{D}\bigl[(p+1)J_2+(n-1)K_2\bigr]
        + e^{2J_0}\!\left(2U^2 + \frac{4n(p+1)}{D}UV + \frac{2n^2(p+1)^2}{D^2}V^2\right).
\end{equation}

\subsubsection{Decoupling via the Eigenbasis}

Write $J_2 = \tilde{U}+\tilde{V}$ and $K_2 = \tilde{U}+(p+1)\tilde{V}$ as before.
Applying $P^{-1}$ to the system \eqref{J2eq}-\eqref{K2eq} and computing $(p+1)(\text{\ref{J2eq}})-(\text{\ref{K2eq}})$ and $-(\text{\ref{J2eq}})+(\text{\ref{K2eq}})$, dividing each by $p$, one obtains the decoupled inhomogeneous equations
\begin{align}
\label{Utildeeq}
\tilde{U}'' &= 2e^{2J_0}\tilde{U} + e^{2J_0}\sigma_U,\\
\label{Vtildeeq}
\tilde{V}'' &= \frac{2n}{D}e^{2J_0}\tilde{V} + e^{2J_0}\sigma_V,
\end{align}
where the source terms are
\begin{align}
\label{sigmaU}
\sigma_U &= 2U^2 - \frac{2n^2(p+1)}{D^2}V^2,\\
\label{sigmaV}
\sigma_V &= \frac{4n}{D}UV + \frac{2n^2(p+2)}{D^2}V^2.
\end{align}

\noindent\textit{Derivation of $\sigma_U$.}
We compute $(p+1)S_J - S_K$ where
$S_J = 2U^2+\frac{4n}{D}UV+\frac{2n^2}{D^2}V^2$ and
$S_K = 2U^2+\frac{4n(p+1)}{D}UV+\frac{2n^2(p+1)^2}{D^2}V^2$:
\begin{align*}
(p+1)S_J - S_K
&= 2pU^2 + \frac{4n}{D}\bigl[(p+1)-(p+1)\bigr]UV
  + \frac{2n^2}{D^2}\bigl[(p+1)-(p+1)^2\bigr]V^2\\
&= 2pU^2 - \frac{2n^2 p(p+1)}{D^2}V^2,
\end{align*}
and dividing by $p$ yields \eqref{sigmaU}.

\noindent\textit{Derivation of $\sigma_V$.}
We compute $-S_J + S_K$:
\begin{equation*}
-S_J+S_K
= \frac{4np}{D}UV + \frac{2n^2}{D^2}\bigl[(p+1)^2-1\bigr]V^2
= \frac{4np}{D}UV + \frac{2n^2 p(p+2)}{D^2}V^2,
\end{equation*}
and dividing by $p$ yields \eqref{sigmaV}.

\subsubsection{Particular Solutions by Variation of Parameters}

The homogeneous parts of \eqref{Utildeeq}-\eqref{Vtildeeq} are precisely the P\"oschl-Teller equations \eqref{Ueq}-\eqref{Veq} already solved at first order.
Denoting the two homogeneous solutions of \eqref{Ueq} by $h_1(\xi)=\coth \xi$ and $h_2(\xi)=\xi\coth \xi-1$ with Wronskian $W[h_1,h_2]=1$, the particular solution of \eqref{Utildeeq} in the $\xi$-variable is
\begin{equation}
\label{Upart}
\tilde{U}_\mathrm{p}(\xi)
= -h_1(\xi)\int^\xi \frac{h_2(\xi')\,\sigma_U(\xi')}{\sinh^2 \xi'}\,d\xi'
  +h_2(\xi)\int^\xi \frac{h_1(\xi')\,\sigma_U(\xi')}{\sinh^2 \xi'}\,d\xi',
\end{equation}
and similarly for $\tilde{V}_\mathrm{p}$ using the P\"oschl-Teller solutions with parameter $\ell_V$.
The full second-order solution is the sum of \eqref{Upart} and the general homogeneous solution with two new integration constants.

\subsection{Structure of the Perturbative Series}

The pattern established at first and second order persists to all orders: the $O(\epsilon^k)$ equations have the same homogeneous P\"oschl-Teller operators as at first order, with source terms that are degree-$k$ polynomials in the lower-order solutions.
Consequently:
\begin{itemize}
\item The homogeneous solutions at every order are the same two functions $h_1,h_2$ (for the $U$-sector) and $P_{\ell_V}, Q_{\ell_V}$ (for the $V$-sector), contributing two new integration constants per sector per order.
\item Particular solutions are obtained by variation of parameters, involving quadratures of products of P\"oschl-Teller eigenfunctions weighted by $\gamma^2/\sinh^2 \xi$.
\item Because $\ell_V$ is never a non-negative integer for physical $(n,p)$, the $V$-sector solutions cannot be expressed in terms of elementary functions at any order. The non-elementary character of the general solution is therefore not an artifact of the perturbative scheme but reflects a genuine analytic obstruction in the Toda equations \eqref{J1}-\eqref{K1}.
\end{itemize}

\section{Weak Dilaton-Coupling Expansion: \texorpdfstring{$\epsilon=a^2$}{epsilon = a squared}}
\label{sec:epsa2}

Section~\ref{sec:dilaton} showed that the dilatonic deformed Toda equations \eqref{equ7}-\eqref{equ8} do not admit a diagonal solution $J=K$ for $a\neq 0$, because $b_1+b_2\neq b_3+b_4$ once the dilaton is switched on.
In this section we show that this obstruction is controlled entirely by $a^2$, and we exploit this fact to construct the leading correction to the pure-Maxwell $\sinh$ solution of Section~\ref{sec:charged} at small dilaton coupling.
Remarkably, the resulting linear system is precisely the perturbative system \eqref{linearmatrix} of Section~\ref{sec:perturb}, now with a specific, non-vanishing source term; identifying the formal deformation parameter $\epsilon$ of Section~\ref{sec:perturb} with $a^2$ therefore turns the perturbative machinery developed there into an honest weak-coupling expansion of the dilatonic solution.

\subsection{Expansion of the Toda Coefficients}

From \eqref{coefficients}, $\Delta=2n(p+1)+(n+p+1)a^2\equiv \Delta_0+(D+1)a^2$, with $\Delta_0\equiv 2n(p+1)$ and $D\equiv n+p$ as in Section~\ref{sec:perturb}.  A short computation gives, to $O(a^4)$,
\begin{align}
\label{bexpand}
b_1 = b_1^{(0)}(1-\eta), \qquad
b_2 = b_2^{(0)}(1-\eta), \qquad
b_3 = b_3^{(0)}(1-\eta), \qquad
b_4 = b_4^{(0)}(1-\eta) + \delta_4,
\end{align}
where $b_i^{(0)}$ are the $a=0$ values of \eqref{coefficients}--which reproduce exactly the pure-Maxwell coefficients $b_1^{(0)}=(D+1)/D$, $b_2^{(0)}=-1/D$, $b_3^{(0)}=(p+1)/D$, $b_4^{(0)}=(n-1)/D$ appearing in \eqref{J1}-\eqref{K1}--and
\begin{equation}
\label{etadelta}
\eta \equiv \frac{D+1}{2n(p+1)}\,a^2, \qquad
\delta_4 \equiv \frac{a^2}{2(p+1)}.
\end{equation}
Both $\eta$ and $\delta_4$ are $O(a^2)$, with no term linear in $a$: this confirms that the deviation of the dilatonic Toda equations from the pure-Maxwell one is controlled entirely by $\epsilon\equiv a^2$.
This is a direct consequence of the $a\to-a$, $\phi\to-\phi$ symmetry of the action \eqref{action2}: the metric sector, being even under this symmetry, can only depend on $a$ through $a^2$.

\subsection{Leading-Order Equations}

Guided by \eqref{bexpand}, we set $\epsilon=a^2$ in the expansion
\eqref{pertexp} of Section~\ref{sec:perturb},
\begin{equation}
\label{epsexp}
J(r) = J_0(r) + a^2 J_1(r) + O(a^4), \qquad
K(r) = J_0(r) + a^2 K_1(r) + O(a^4),
\end{equation}
with $J_0$ the diagonal background \eqref{background}.
Expanding \eqref{equ7}-\eqref{equ8} to $O(a^2)$ using \eqref{bexpand}-\eqref{etadelta} and the identities $b_1^{(0)}+b_2^{(0)}=b_3^{(0)}+b_4^{(0)}=1$ already used in Section~\ref{sec:dilaton} gives
\begin{align}
\label{J1source}
J_1'' &= \frac{2e^{2J_0}}{D}\bigl[(D+1)J_1-K_1\bigr]
       \;-\; \frac{D+1}{n(p+1)}\,J_0\,e^{2J_0},\\
\label{K1source}
K_1'' &= \frac{2e^{2J_0}}{D}\bigl[(p+1)J_1+(n-1)K_1\bigr]
       \;-\; \frac{1}{n}\,J_0\,e^{2J_0}.
\end{align}
The homogeneous (matrix) part of \eqref{J1source}-\eqref{K1source} is identical to \eqref{linearmatrix}: it originates from linearizing $e^{2b_1^{(0)}J+2b_2^{(0)}K}$ in $(J_1,K_1)$ and is therefore insensitive to the physical origin of the perturbation.
What is new here is the source term on the right, proportional to $J_0\,e^{2J_0}$: it originates from the $O(a^2)$ shift of the coefficients $b_i$ themselves, evaluated on the diagonal background.
Physically, it encodes the fact that the diagonal solution back-reacts on itself once the exponential dilaton coupling is switched on, even though at $a=0$ it solves the undeformed Toda equations exactly.

\subsection{Decoupling: Sourced P\"oschl-Teller Equations}

We diagonalize \eqref{J1source}-\eqref{K1source} exactly as in Section~\ref{sec:perturb}, writing $J_1=U+V$, $K_1=U+(p+1)V$ in terms of the eigenvectors of $\mathbf{M}$ found there.
Because the transformation \eqref{UVdef}-\eqref{UVinv} is linear, it decouples the source term in the same way it decoupled the homogeneous matrix, and one finds
\begin{align}
\label{Usourced}
U'' &= 2\,e^{2J_0}\,U \;-\; \frac{D}{np}\,J_0\,e^{2J_0},\\
\label{Vsourced}
V'' &= \frac{2n}{D}\,e^{2J_0}\,V \;+\; \frac{1}{p(p+1)}\,J_0\,e^{2J_0}.
\end{align}
These are inhomogeneous P\"oschl-Teller equations with exactly the same $\ell$-values found in Section~\ref{sec:perturb}: $\ell=1$ for $U$ and $\ell=\ell_V$ given by \eqref{ellV} for $V$.
The homogeneous solutions are therefore again $h_1(\xi)=\coth \xi$, $h_2(\xi)=\xi\coth \xi-1$ for $U$, and $P_{\ell_V}(\coth \xi)$, $Q_{\ell_V}(\coth \xi)$ for $V$, with $\xi=\gamma r+\delta$.

\subsection{The \texorpdfstring{$U$}{U}-Sector: A Closed-Form Elementary Solution}

Using $J_0(\xi)=\ln\gamma-\ln\sinh \xi$ and $e^{2J_0}=\gamma^2/\sinh^2 \xi$, equation \eqref{Usourced} becomes, in the $\xi$-variable,
\begin{equation}
\label{Us}
\frac{d^2U}{d\xi^2} = \frac{2}{\sinh^2 \xi}\,U - \frac{D}{np}\,\frac{J_0(\xi)}{\sinh^2 \xi}.
\end{equation}
Variation of parameters (Wronskian $W[h_1,h_2]=1$) gives the particular solution
\begin{equation}
\label{Upartsourced}
U_\mathrm{p}(\xi) = \frac{D}{np}\left[
  h_1(\xi)\int^\xi \frac{h_2(\xi')\,J_0(\xi')}{\sinh^2 \xi'}\,d\xi'
  - h_2(\xi)\int^\xi \frac{h_1(\xi')\,J_0(\xi')}{\sinh^2 \xi'}\,d\xi'
\right].
\end{equation}
The $h_1$-weighted quadrature is elementary: using $J_0'(\xi)=-\coth \xi$ and integrating by parts twice,
\begin{equation}
\label{h1integral}
\int \frac{\coth \xi\,J_0(\xi)}{\sinh^2 \xi}\,d\xi
= -\frac{J_0(\xi)}{2\sinh^2 \xi} + \frac{\coth^2 \xi}{4} + \text{const}.
\end{equation}
The $h_2$-weighted quadrature is elementary as well.
The essential point is that the source in \eqref{Upartsourced} is weighted by $\mathrm{csch}^2 \xi$: using the homogeneous equation $h_2''=2\,h_2/\sinh^2 \xi$ to write $h_2/\sinh^2 \xi=\tfrac12 h_2''$ and integrating by parts twice, the quadrature reduces to $\int J_0(\xi)\,\mathrm{csch}^2 \xi\,d\xi=-J_0\coth \xi-\xi+\coth \xi$, which is elementary--\emph{not} to the bare integral $\int J_0(\xi)\,d\xi$, whose
evaluation would require the dilogarithm.
Explicitly,
\begin{equation}
\label{h2integral}
\int \frac{h_2(\xi)\,J_0(\xi)}{\sinh^2 \xi}\,d\xi
= -\frac{\xi\,J_0(\xi)}{2\sinh^2 \xi}+\frac{J_0(\xi)\coth \xi}{2}
  +\frac{\xi}{2}-\frac{\coth \xi}{4}+\frac{\xi}{4\sinh^2 \xi}+\text{const}.
\end{equation}
Substituting \eqref{h1integral}-\eqref{h2integral} into \eqref{Upartsourced}, the entire $U$-sector particular solution collapses to the remarkably compact closed form
\begin{equation}
\label{Upclosed}
U_\mathrm{p}(\xi) = \frac{D}{4np}\Bigl(\xi\coth \xi + 2\,J_0(\xi)\Bigr),
\end{equation}
as is verified directly by substitution into \eqref{Us}.  Thus the $\ell=1$ ($U$) sector remains \emph{fully elementary} even after it is sourced by the exponential dilaton coupling; despite the
$\mathrm{csch}^2 \xi$ weighting of the source, no dilogarithm appears.
As we now show, the same $\mathrm{csch}^2 \xi$ mechanism renders the $V$-sector particular solution elementary as well, so that the genuine obstruction resides not in the particular solution but in the homogeneous modes.

\subsection{The \texorpdfstring{$V$}{V}-Sector}

As in Section~\ref{sec:perturb}, since $\ell_V$ is not a non-negative integer for any physical $(n,p)$, the \emph{homogeneous} solutions of \eqref{Vsourced} are the genuinely non-elementary Legendre functions $P_{\ell_V}(\coth \xi),Q_{\ell_V}(\coth \xi)$.
The \emph{particular} solution, by contrast, is elementary: its source $+\tfrac{1}{p(p+1)}J_0\,e^{2J_0}$ carries the same $\mathrm{csch}^2 \xi$ weighting as the $U$-sector source in \eqref{Us}, so the integration-by-parts mechanism that produced \eqref{Upclosed} applies unchanged and no Legendre quadrature is required.
Substituting the ansatz $V=\kappa_V\ln(\sinh \xi) + \text{const}$ into \eqref{Vsourced} and matching the coefficient of $J_0/\sinh^2 \xi$ fixes $\kappa_V=D/[2np(p+1)]$, giving the compact closed form
\begin{equation}
\label{Vpclosed}
V_\mathrm{p}(\xi) = -\frac{D}{2np(p+1)}\Bigl(J_0(\xi)+\frac{D}{2n}\Bigr)
= \frac{D}{2np(p+1)}\,\ln(\sinh \xi) + \text{const},
\end{equation}
as is verified directly by substitution into \eqref{Vsourced}.
This is the perturbative image of the closed-form, distinct-power $\sinh$ branch anticipated in Section~\ref{sec:dilaton} [cf.\ Eq.~\eqref{dilsol} and the following discussion]: its exponents $m_K=1$, $m_J=1+(n+p)a^2/[2n(p+1)]$ reduce at $O(a^2)$ to exactly the elementary $U$-$V$ pair \eqref{Upclosed}, \eqref{Vpclosed}, through $J_1=U+V$, $K_1=U+(p+1)V$.

The genuine non-elementary obstruction therefore does \emph{not} reside in the particular solution, contrary to a naive extrapolation from Section~\ref{sec:perturb}.
It reappears instead at the level of \emph{horizon regularity}.
Like $U_\mathrm{p}$, the particular solution \eqref{Vpclosed} grows linearly as $\xi\to\infty$; but whereas the linear growth of $U_\mathrm{p}$ is removed by the \emph{pure-gauge} homogeneous
mode $h_2$ [Eq.~\eqref{Uinfclosed}], the growth of $V_\mathrm{p}$ can be cancelled only by the \emph{non-elementary} mode $Q_{\ell_V}(\coth \xi)$, which is not generated by any residual coordinate freedom.
A finite-area Killing horizon therefore forces a non-zero admixture of $Q_{\ell_V}$, so it is this regularity-selected homogeneous mode--not the particular solution--that carries the genuine non-elementary content of the dilatonic $V$-sector.
The fixing of the $V$-sector integration constants by horizon regularity is analyzed in Section~\ref{subsec:thermoeps_hair}.

\subsection{Leading-Order Dilaton Profile}

The physically most transparent result of this section does not require solving \eqref{Usourced}-\eqref{Vsourced} at all: it follows directly from the linear equation \eqref{EOM6sol}.
Since the dilaton vanishes identically at $a=0$, the natural boundary condition on the linear zero mode is $\kappa_\phi=\mu_\phi=0$, and \eqref{EOM6sol} gives, to leading order in $a$,
\begin{equation}
\label{phileading}
\phi(r) = a\bigl[A_0(r)-C_0\bigr] + O(a^3),
\end{equation}
where $A_0(r)$ and $C_0$ are the pure-Maxwell background functions \eqref{Aexplicit} and \eqref{C0value}.
Because the metric functions $A,F,C$ receive corrections only at $O(a^2)$--through $J_1,K_1$ above--while $\phi$ is $O(a)$, the leading-order dilaton profile \eqref{phileading} is simply the pure-Maxwell metric function $A_0$, offset by the constant torus modulus $C_0$.
This odd-in-$a$, even-in-$a$ split between $\phi$ and $(A,F,C)$ is exactly the pattern required by the $a\to-a$, $\phi\to-\phi$ symmetry noted above.

Equation \eqref{phileading} makes the qualitative statement of Section~\ref{sec:dilaton}--that the dilatonic branch has ``a non-trivial dilaton profile''--fully explicit at leading order.
Near the horizon ($r\to\infty$), $A_0(r)\to-\gamma r+\text{const}$, so $\phi$ grows linearly, $\phi\to -a\gamma r$, the characteristic linear-dilaton behavior of charged dilatonic black branes.
Near the AdS boundary ($r\to 0$), $J_0(r)\to+\infty$ logarithmically and hence so does $A_0$, so that $\phi$ diverges at the conformal boundary as well. Concretely, in the Fefferman-Graham coordinate $\zeta=r^{1/(p+1)}$ this profile behaves as $\phi=-a\ln\zeta+\text{const}+O(\zeta^{\,p+1})$: the leading divergence is a pure logarithm whose $\gamma$-independent coefficient $\phi_{(0)}=-a$, fixed by the exponential coupling $e^{a\phi}$ in \eqref{action2} alone, is the non-normalizable source of a marginal (dimension-$(p+1)$) operator--the marginal coupling of the dual CFT--while the $\gamma$-dependent normalizable response first enters at the subleading order $\zeta^{\,p+1}$. This holographic reading is made quantitative in Section~\ref{subsec:thermoeps_firstlaw} [Eq.~\eqref{phisource}], where the constancy of $\phi_{(0)}$ along the family is what removes the dilaton work term from the first law.

\subsection{Summary}

To summarize, at leading order in $\epsilon=a^2$ the dilatonic charged brane is characterized by:
\begin{itemize}
\item a metric identical to the pure-Maxwell solution of Section~\ref{sec:charged}, corrected at $O(a^2)$ by the sourced P\"oschl-Teller system \eqref{Usourced}-\eqref{Vsourced}, whose $U$-sector and $V$-sector particular solution are elementary and given in closed form by \eqref{Upclosed} and \eqref{Vpclosed}; the only non-elementary structure is the pair of homogeneous modes $P_{\ell_V},Q_{\ell_V}$, fixed by horizon regularity ($Q_{\ell_V}$) and AdS boundary normalizability ($P_{\ell_V}=0$), so that both $V$-sector constants are determined in terms of $(n,p,q,\gamma)$ with no free parameter left over (Section~\ref{subsec:thermoeps_hair});
\item a dilaton profile that is $O(a)$ and given in closed form by \eqref{phileading}.
\end{itemize}
This provides the leading term of the perturbative construction anticipated at the end of Section~\ref{sec:dilaton}; the associated correction to the Hawking temperature and entropy density is worked out in Section~\ref{sec:thermo_eps}, while the general fully non-linear dilatonic branch, and higher orders in $a^2$, are left for future work.

\section{Thermodynamics of the Exact Solution}
\label{sec:thermo}

In this section we compute the Hawking temperature and the Bekenstein-Hawking entropy density of the special exact solution constructed in Section~\ref{sec:charged}.

\subsection{Metric Functions at the Horizon}
\label{subsec:horizon}

Recall the exact solution with $\alpha=\gamma$, $\beta=\delta=0$:
\begin{equation}
e^{-J_0(r)} = \frac{\sinh(\gamma r)}{\gamma}.
\end{equation}
On this diagonal solution $J=K$ implies, via \eqref{J2}-\eqref{K2}, that the torus function $C(r)$ is constant,
\begin{equation}
\label{C0value}
C_0 = \frac{c_K-c_J}{n+p}
   = \frac{1}{2n}\ln\!\frac{(p+1)\,q^2\,l^2}{2(n+p)(n+p+1)},
\end{equation}
while the metric functions $A_0$ and $F_0$ are
\begin{align}
\label{Aexplicit}
A_0(r) &= \frac{J_0(r) + \mu}{p+1} - \frac{p\,\gamma\,r}{p+1},\\
\label{Fexplicit}
F_0(r) &= \frac{J_0(r) + \mu}{p+1} + \frac{\gamma\,r}{p+1},
\end{align}
where $\mu \equiv -c_J - (n-1)C_0$.  The gauge function $B_0 = J_0 + \mu + nC_0$ is a function of $r$ only through $J_0$.

The horizon is located at $r\to\infty$, where $e^{J_0}\to 2\gamma\,e^{-\gamma r}\to 0$ so that $e^{2A_0}\to 0$.
Using $J_0(r)\approx \ln(2\gamma)-\gamma r$ for $r\to\infty$, one finds
\begin{equation}
A_0(r)\;\xrightarrow{r\to\infty}\;-\gamma\,r + \text{const},\qquad
B_0(r)\;\xrightarrow{r\to\infty}\;-\gamma\,r + \text{const},
\end{equation}
so both $g_{tt}=-e^{2A_0}$ and $g_{rr}=e^{2B_0}$ vanish exponentially at the horizon, consistently with the Killing horizon structure.

\subsection{Hawking Temperature}
\label{subsec:hawking}

For a static metric $ds^2=-e^{2A}dt^2+e^{2B}dr^2+\cdots$, the surface gravity of the Killing horizon $\xi^\mu=(\partial_t)^\mu$ is given by
\cite{Wald:1984rg,Townsend:1997kt}
\begin{equation}
\label{surfacegravity}
\kappa = \lim_{r\to r_H}|A'(r)|\,e^{A(r)-B(r)}.
\end{equation}
The Hawking temperature is $T_H = \kappa/(2\pi)$.

We evaluate \eqref{surfacegravity} at $r\to\infty$.
From~\eqref{Aexplicit}:
\begin{equation}
A_0'(r) = \frac{J_0'(r)}{p+1} - \frac{p\,\gamma}{p+1},
\qquad
J_0'(r) = -\gamma\coth(\gamma r)\;\xrightarrow{r\to\infty}\;-\gamma.
\end{equation}
Therefore $A_0'\to -\gamma(p+1)/(p+1) = -\gamma$ as $r\to\infty$, giving $|A_0'|=\gamma$.

For $A_0-B_0$ at $r\to\infty$:
\begin{align}
A_0 - B_0
&= \frac{J_0+\mu-p\gamma r}{p+1} - (J_0+\mu+nC_0)\notag\\
&= -\frac{p}{p+1}J_0 - \frac{p}{p+1}\mu - \frac{p\gamma r}{p+1} - nC_0.
\end{align}
Substituting $J_0\approx\ln(2\gamma)-\gamma r$, the $\gamma r$ terms cancel:
\begin{equation}
\label{ABinf}
(A_0-B_0)\big|_{r\to\infty}
= -\frac{p}{p+1}\ln(2\gamma) - \frac{p}{p+1}\mu - nC_0
\;\equiv\; \Omega,
\end{equation}
which is a pure constant (independent of $r$).  Substituting $\mu = -c_J-(n-1)C_0$ and the explicit forms of $c_J$, $C_0$:
\begin{equation}
\label{Omegaexplicit}
\Omega = -\frac{p}{p+1}\ln(2\gamma)
         -\frac{1}{2(p+1)}\ln\!\frac{(p+1)q^2}{2}
         +\frac{1}{2}\ln\!\frac{(n+p)(n+p+1)}{l^2}.
\end{equation}
The surface gravity is therefore
\begin{equation}
\label{kappafinal}
\kappa = \gamma\,e^{\Omega},
\end{equation}
and the Hawking temperature is
\begin{equation}
\label{THfinal}
T_H = \frac{\gamma\,e^{\Omega}}{2\pi}
    = \frac{\gamma^{\frac{1}{p+1}}}{2\pi}
      \cdot\frac{1}{2^{\frac{p}{p+1}}}
      \cdot\left[\frac{(n+p)(n+p+1)}{l^2}\right]^{\!\frac{1}{2}}
      \cdot\left[\frac{(p+1)q^2}{2}\right]^{\!-\frac{1}{2(p+1)}}.
\end{equation}
For the simplest non-trivial case $n=p=1$, substituting $(n+p)(n+p+1)=6$ and $p+1=2$ into the formula gives
\begin{equation}
T_H\big|_{n=p=1}
= \frac{\sqrt{3\gamma}}{2\pi\,l\,\sqrt{q}}.
\end{equation}

Several features of \eqref{THfinal} deserve comment.
First, $T_H$ depends only on the solution parameters $\gamma$, $q$, and the fixed AdS radius $l$; it does not depend on the coordinate parameter $r_+$ introduced in the change of variables \eqref{coordtrans}, confirming coordinate independence.
Second, the temperature vanishes as $\gamma\to 0$ and diverges as $q\to 0$.
The limit $\gamma\to 0$ corresponds to the near-extremal limit, while the $q\to 0$ limit removes the magnetic charge and is outside the validity of the present ansatz.
Third, $T_H\propto \gamma^{1/(p+1)}$ exactly, reflecting the $(p+1)$-dimensional nature of the brane worldvolume.

\subsection{Bekenstein-Hawking Entropy Density}
\label{subsec:entropy}

The Bekenstein-Hawking entropy is $S=A_\mathrm{hor}/(4G_{n+p+2})$ \cite{Bekenstein:1973ur,Hawking:1975vcx}.
Since the worldvolume directions $z^i$ are non-compact, we work with the entropy density $s = S/V_p$, where $V_p=\prod_i L_{z_i}$ is the regulated worldvolume volume.

At the horizon ($r\to\infty$) the induced metric on a constant-$t$, constant-$r$ slice has components
\begin{equation}
ds^2_\mathrm{hor}
= e^{2F_\mathrm{hor}}\sum_{i=1}^p dz_i^2
+ e^{2C_0}\sum_{i=1}^n d\theta_i^2,
\end{equation}
where $F_\mathrm{hor}\equiv\lim_{r\to\infty}F_0(r)$.  From~\eqref{Fexplicit} and $J_0\approx\ln(2\gamma)-\gamma r$:
\begin{equation}
\label{Fhor}
F_\mathrm{hor}
= \lim_{r\to\infty}\frac{J_0+\mu+\gamma r}{p+1}
= \frac{\ln(2\gamma)+\mu}{p+1}.
\end{equation}
The $(n+p)$-dimensional horizon area density (per unit $V_p\cdot(2\pi)^n$) is
\begin{equation}
\frac{A_\mathrm{hor}}{V_p(2\pi)^n}
= e^{p\,F_\mathrm{hor}+n\,C_0}.
\end{equation}
Substituting \eqref{Fhor} and \eqref{C0value}:
\begin{equation}
\label{sfinal}
s = \frac{(2\pi)^n\,e^{p\,F_\mathrm{hor}+n\,C_0}}{4G_{n+p+2}}
  = \frac{(2\pi)^n}{4G_{n+p+2}}
    \cdot(2\gamma)^{\frac{p}{p+1}}
    \cdot\left[\frac{(p+1)q^2}{2}\right]^{\!\frac{1}{2(p+1)}}
    \cdot\left[\frac{l^2}{(n+p)(n+p+1)}\right]^{\!\frac{1}{2}}.
\end{equation}
For $n=p=1$, substituting $p+1=2$, $(n+p)(n+p+1)=6$ into \eqref{sfinal}:
\begin{equation}
s\big|_{n=p=1}
= \frac{\pi\,l\,\sqrt{\gamma\,q}}{2\sqrt{3}\,G_4}.
\end{equation}

Note that $T_H$ and $s$ depend on $\gamma$ in complementary ways: $T_H\propto\gamma^{1/(p+1)}$ while $s\propto\gamma^{p/(p+1)}$, so that the product
\begin{equation}
T_H \cdot s \propto \gamma
\end{equation}
is linear in the single non-charge parameter $\gamma$.
If the first law $d\mathcal{M}=T_H\,ds$ holds for this one-parameter family, integrating with the above power laws predicts a mass density $\mathcal{M}\propto\gamma$ and hence a Smarr-type relation $T_H\,s\propto\mathcal{M}$.
In the next subsection we compute $\mathcal{M}$ independently by holographic renormalization and confirm that the first law indeed holds.

\subsection{Holographic Mass and the First Law}
\label{subsec:holomass}

We now compute the mass density of the brane by holographic renormalization and verify the first law directly.
Near the conformal boundary $r\to0$ one has $J_0=-\ln r-\tfrac16\gamma^2r^2+O(r^4)$, so the metric functions \eqref{Aexplicit}-\eqref{Fexplicit} expand as
\begin{align}
e^{2A_0} &= e^{2\mu/(p+1)}\,r^{-2/(p+1)}
         \Bigl[\,1-\tfrac{2p\gamma}{p+1}\,r+O(r^2)\Bigr],\\
e^{2F_0} &= e^{2\mu/(p+1)}\,r^{-2/(p+1)}
         \Bigl[\,1+\tfrac{2\gamma}{p+1}\,r+O(r^2)\Bigr],
\end{align}
while $e^{2C}=e^{2C_0}$ is constant and $e^{2B_0}=e^{2(\mu+nC_0)}r^{-2}\,[1+O(r^2)]$.
The Fefferman-Graham coordinate $\zeta=r^{1/(p+1)}$ brings the $(t,z_i,r)$ block to the standard form
\begin{equation}
ds^2_{p+2}=\frac{L^2}{\zeta^2}\bigl(d\zeta^2+g_{ij}\,dx^i dx^j\bigr),
\qquad L=(p+1)\,e^{\mu+nC_0},\qquad x^i=(t,z_1,\dots,z_p),
\end{equation}
with a flat boundary metric $g^{(0)}_{ij}=c_0\,\eta_{ij}$ and $c_0=e^{2\mu/(p+1)}/L^2$; the torus $T^n$ factors off as an internal space of fixed volume.
Since the first subleading correction enters at order $\zeta^{p+1}$--the intermediate coefficients $g^{(2)},\dots,g^{(p)}$ vanishing, exactly as for a planar AdS-Schwarzschild hole--the
Fefferman-Graham data reduce to
\begin{equation}
g^{(0)}_{ij}=c_0\,\eta_{ij},\qquad
g^{(p+1)}_{tt}=\frac{2p\gamma}{p+1}\,c_0,\qquad
g^{(p+1)}_{z_iz_i}=\frac{2\gamma}{p+1}\,c_0 .
\end{equation}

Because the modulus $C\equiv C_0$ is constant throughout the solution, reduction on $T^n$ is exact and merely rescales Newton's constant, $G_{p+2}=G_{n+p+2}/[(2\pi)^n e^{nC_0}]$, leaving an asymptotically AdS$_{p+2}$ Einstein theory of radius $L$.
The renormalized boundary stress tensor is then \cite{deHaro:2000vlm}
\begin{equation}
\label{bdystress}
\langle T_{ij}\rangle
=\frac{(p+1)\,L^{p}}{16\pi G_{p+2}}\,g^{(p+1)}_{ij},
\end{equation}
with no anomalous piece, since $g^{(0)}$ is flat.
The energy density conjugate to $\partial_t$, per unit coordinate worldvolume $d^pz$, is $\mathcal{M}=c_0^{(p-1)/2}\,\langle T_{tt}\rangle$, the factor $c_0^{(p-1)/2}$ accounting for the non-unit normalization of the boundary metric.
Using $c_0^{(p+1)/2}L^{p}=e^{\mu}/L$ and $e^{\mu}e^{nC_0}/L=1/(p+1)$, all dependence on $\mu$ and $C_0$--and hence on the charge $q$ and the AdS radius $l$--cancels, leaving
\begin{equation}
\label{massfinal}
\mathcal{M}=\frac{p\,\gamma\,(2\pi)^n}{8\pi\,(p+1)\,G_{n+p+2}} .
\end{equation}

The result \eqref{massfinal} passes two independent checks.
First, combining \eqref{THfinal} and \eqref{sfinal} yields the compact relation
\begin{equation}
\label{THscompact}
T_H\,s=\frac{\gamma\,(2\pi)^n}{8\pi\,G_{n+p+2}},
\end{equation}
in which the redshift factor $e^{\Omega}$ of \eqref{Omegaexplicit} cancels between temperature and entropy, so that
\begin{equation}
\label{firstlawM}
\mathcal{M}=\frac{p}{p+1}\,T_H\,s .
\end{equation}
Equation \eqref{firstlawM} is exactly the integrated first law $d\mathcal{M}=T_H\,ds$ for the one-parameter family: with $s\propto\gamma^{p/(p+1)}$ and $T_H\propto\gamma^{1/(p+1)}$ one has
$\int T_H\,ds=\tfrac{p}{p+1}T_H\,s$.
Second, the spatial components of \eqref{bdystress} give a pressure $P=\mathcal{M}/p$, so that $-\mathcal{M}+pP=0$: the boundary stress tensor is traceless, as required of a $(p+1)$-dimensional conformal fluid.
These results confirm the Smarr-type scaling $\mathcal{M}\propto T_H\,s\propto\gamma$ anticipated in Section~\ref{subsec:entropy} and promote the entropy law $s\propto T_H^{p}$ to a complete thermodynamic description at the level of the boundary stress tensor.

\subsection{Free Energy and Conformal Equation of State}
\label{subsec:freeenergy}

With the pressure $P=\mathcal{M}/p$ read off from \eqref{bdystress} and the relation \eqref{firstlawM}, the Helmholtz free-energy density of the brane follows at once,
\begin{equation}
\label{freeenergy}
f \;\equiv\; \mathcal{M}-T_H\,s
\;=\; -\frac{1}{p+1}\,T_H\,s
\;=\; -\frac{\gamma\,(2\pi)^n}{8\pi(p+1)\,G_{n+p+2}}
\;=\; -\frac{\mathcal{M}}{p}\;=\;-P .
\end{equation}
Writing $\varepsilon\equiv\mathcal{M}$ for the energy density, these are the conformal equation of state of the dual $(p+1)$-dimensional fluid,
\begin{equation}
\label{eos}
\varepsilon = p\,P,\qquad
T_H\,s=(p+1)\,P=\varepsilon+P,\qquad
f=-P,
\end{equation}
the first relation being the tracelessness $-\varepsilon+pP=0$ of the boundary stress tensor established above.  Since $T_H\propto\gamma^{1/(p+1)}$ at fixed $(q,l)$, \eqref{freeenergy} is the Stefan-Boltzmann law $f=-c(q,l)\,T_H^{\,p+1}$, whose coefficient depends on the magnetic charge and the AdS radius but not on the temperature scale $\gamma$.

Equivalently, the first two relations in \eqref{eos} combine into the Euler relation
\begin{equation}
\label{euler}
\varepsilon = T_H\,s - P ,
\end{equation}
which is precisely the mass formula $\mathcal{M}=\tfrac{p}{p+1}T_H s$ of \eqref{firstlawM} rewritten.
See also \cite{Yun:2026}.
Because the solution family is one-parameter--labelled by the non-extremality scale $\gamma$ at fixed magnetic charge $q$--the variation defining the first law holds $q$ fixed ($dq=0$), so $d\mathcal{M}=T_H\,ds$ follows with no $\Phi\,dq$ term by construction.
Notably, \eqref{euler} contains no chemical-potential term.
The magnetic charge does enter the temperature and entropy separately--indeed $s\propto q^{1/(p+1)}$--but only as a fixed background scale, setting the coefficient $c(q,l)$ of the Stefan-Boltzmann law rather than acting as a thermodynamic charge conjugate to a potential.
(Had one instead posited a work term $\Phi\,dq$ with $\Phi=-T_H\,\partial s/\partial q$, the resulting $\Phi\,q=-\tfrac{1}{p+1}T_H s$ would spoil \eqref{euler}; since it was established independently from the boundary stress tensor, no such term is present along this one-parameter family.
A first-principles covariant-phase-space (Wald) derivation confirming this--or fixing the form of any magnetic work term when $q$ is varied--is left to future work.)
The equilibrium first law is therefore $d\mathcal{M}=T_H\,ds$ at fixed magnetic charge, with no additional $\Phi\,dq$ contribution.

At $O(a^2)$ the dilaton develops scalar hair (Section~\ref{subsec:thermoeps_hair}), but this hair does \emph{not} modify the first law along the present family: unlike the magnetically charged AdS black holes of~\cite{Lu:2013ura}, whose first law carries a scalar-charge work term, here the dilaton source is frozen by the coupling and $d\mathcal{M}=T_H\,ds$ persists at $O(a^2)$ (Section~\ref{subsec:thermoeps_firstlaw}).

\subsection{Free Energy from the Renormalized On-Shell Action}
\label{subsec:onshellaction}

The free-energy density \eqref{freeenergy} was inferred thermodynamically, from $\mathcal{M}$ and $T_H s$.
We now derive it directly from the renormalized Euclidean on-shell action $I_E=\beta V_p\,f$, an independent first-principles check.
As established around \eqref{bdystress}, the constancy of the modulus $C\equiv C_0$ makes the reduction on $T^n$ exact: the $(t,z_i,r)$ block is precisely a planar Einstein-AdS$_{p+2}$ black hole of radius $L=(p+1)e^{\mu+nC_0}$ with Newton constant $G_{p+2}=G_{n+p+2}/[(2\pi)^n e^{nC_0}]$, the magnetic flux and the bulk cosmological term \eqref{Lambda} combining into the single effective constant $\Lambda_{p+2}=-p(p+1)/(2L^2)$.
Its Ricci scalar is accordingly the constant $\mathcal{R}=-(p+1)(p+2)/L^2$, so on shell $\mathcal{R}-2\Lambda_{p+2}=-2(p+1)/L^2$.
The renormalized action, with Gibbons-Hawking-York term and the leading counterterm for a flat boundary \cite{deHaro:2000vlm}, is
\begin{equation}
\label{Irenaction}
\begin{aligned}
I_E={}&-\frac{1}{16\pi G_{p+2}}\int d^{p+2}x\,\sqrt{g}\,(\mathcal{R}-2\Lambda_{p+2}) -\frac{1}{8\pi G_{p+2}}\int d^{p+1}x\,\sqrt{h}\,K\\
     &+\frac{p}{8\pi G_{p+2}\,L}\int d^{p+1}x\,\sqrt{h}.
\end{aligned}
\end{equation}
In the radial coordinate of \eqref{metricansatz} (boundary at $r\to0$, horizon at $r\to\infty$) the reduced volume element and boundary data are
\begin{equation}
\begin{aligned}
\sqrt{g}&=e^{A_0+B_0+pF_0}=e^{2\mu+nC_0}\,\frac{\gamma^2}{\sinh^2\gamma r},
\qquad
\sqrt{h}=e^{A_0+pF_0}=\frac{e^{\mu}\gamma}{\sinh\gamma r},\\
K&=-e^{-B_0}J_0'=e^{-\mu-nC_0}\cosh\gamma r,
\end{aligned}
\end{equation}
the last for the outward-pointing normal, with $J_0'=-\gamma\coth\gamma r$.
Cutting the radial integral off at $r=r_c$, using $\int_{r_c}^{\infty}\gamma^2\,\mathrm{csch}^2(\gamma r)\,dr =\gamma[\coth(\gamma r_c)-1]$, and inserting $L^2=(p+1)^2 e^{2\mu+2nC_0}$--which removes all dependence on $\mu$--the three terms of \eqref{Irenaction} combine into
\begin{equation}
\label{onshellthree}
\begin{aligned}
I_E&=\frac{\beta V_p\,e^{-nC_0}\gamma}{8\pi(p+1)G_{p+2}}
\Bigl[\,\underbrace{(\coth\gamma r_c-1)}_{\text{bulk}} \underbrace{-(p+1)\coth\gamma r_c}_{\text{GHY}} + \underbrace{\tfrac{p}{\sinh\gamma r_c}}_{\text{c.t.}}\,\Bigr]\\
   &=\frac{\beta V_p\,e^{-nC_0}\gamma}{8\pi(p+1)G_{p+2}}
 \Bigl[-p\tanh\tfrac{\gamma r_c}{2}-1\Bigr],
\end{aligned}
\end{equation}
where the $1/r_c$ divergences of the bulk and Gibbons-Hawking terms cancel against the counterterm.
Removing the cutoff and restoring $G_{p+2}$,
\begin{equation}
\label{fonshell}
f=\lim_{r_c\to0}\frac{I_E}{\beta V_p}
 =-\frac{e^{-nC_0}\gamma}{8\pi(p+1)G_{p+2}}
 =-\frac{\gamma\,(2\pi)^n}{8\pi(p+1)\,G_{n+p+2}},
\end{equation}
in which the factor $e^{-nC_0}$ cancels.
This is exactly the thermodynamic free energy \eqref{freeenergy}: as with the holographic mass \eqref{massfinal}, all dependence on $\mu$ and $C_0$--and hence on the charge $q$ and the AdS
radius $l$--drops out, and the renormalized on-shell action reproduces $f=\mathcal{M}-T_H s=-P$ from first principles.

\subsection{Holographic Interpretation}
\label{subsec:holointerpret}

Via AdS/CFT \cite{Maldacena:1997re,Witten:1998qj,Aharony:1999ti}, the black brane solutions constructed here are dual to thermal states of a
strongly coupled CFT living on the boundary spacetime $\mathbb{R}^{p,1}\times T^n$.
In this correspondence the Hawking temperature $T_H$ is identified with the temperature of the dual thermal state, and the entropy density $s$ with the thermodynamic entropy density of the CFT.

The UV completion of the dual description is the conformal field theory dual to the asymptotically AdS region.  For a CFT in $d+1$ spacetime dimensions with flat (toroidal) spatial topology, the Stefan-Boltzmann law gives $s\sim T^d$ \cite{Witten:1998zw}.
Here the boundary is $(p+1+n)$-dimensional, but the exact solution has a \emph{fixed} torus modulus $C_0$, so the torus volume $(2\pi)^n e^{nC_0}$ enters the entropy density only as an overall constant and introduces no additional thermal (temperature-dependent) directions.
Equivalently, the Kaluza-Klein tower on $T^n$ is frozen out rather than contributing continuous modes; the thermal scaling is therefore that of the $(p+1)$-dimensional worldvolume theory, so one expects $s\sim T_H^p$, and this scaling holds for the entire solution family.
From \eqref{THfinal} and \eqref{sfinal} one reads off
\begin{equation}
T_H \propto \gamma^{\frac{1}{p+1}}\,q^{-\frac{1}{p+1}},
\qquad
s \propto \gamma^{\frac{p}{p+1}}\,q^{\frac{1}{p+1}}.
\end{equation}
Eliminating $\gamma$ via $\gamma\propto T_H^{p+1}\,q$ and substituting into
$s$ gives
\begin{equation}
\frac{s}{T_H^p} \propto q\,l^{\,p+1},
\end{equation}
which is independent of $\gamma$, so $s\propto T_H^p$ holds along curves of fixed $(q,l)$.
As an explicit check, at $n=p=1$
\begin{equation}
\frac{s}{T_H}\bigg|_{n=p=1}
= \frac{\pi^2 l^2 q}{3G_4},
\end{equation}
which is indeed $\gamma$-independent.  Two caveats qualify this interpretation.
First, the magnetic charge introduces a scale and breaks conformal invariance, so the dual is not exactly a CFT; the survival of the Stefan-Boltzmann scaling $s\sim T_H^p$ indicates only that the leading thermal entropy retains the effective $(p+1)$-dimensional character inherited from the UV CFT.
Second, the coefficient $s/T_H^p\propto q\,l^{\,p+1}$ depends on the magnetic charge rather than being a fixed pure number, so it represents a charge-dependent effective number of thermal degrees of freedom rather than a universal central charge; a sharper identification is provided by the free energy from holographic renormalization, computed in Section~\ref{subsec:onshellaction}: it confirms $f=-P$ with precisely this charge-dependent coefficient, so the non-universality is a genuine consequence of the charge scale breaking conformal invariance rather than an artifact of the thermodynamic derivation.

\section{Thermodynamics at Weak Dilaton Coupling (\texorpdfstring{$\epsilon=a^2$}{epsilon = a squared})}
\label{sec:thermo_eps}

We now extend the thermodynamic analysis of Section~\ref{sec:thermo} to the weakly coupled dilatonic brane constructed perturbatively in Section~\ref{sec:epsa2}.
Because $J(r)$ and $K(r)$ no longer coincide once $a\neq0$, the horizon geometry is not literally the diagonal solution of Section~\ref{sec:charged}; nevertheless the general definitions of surface gravity and horizon area apply unchanged, and we use them to determine how $T_H$ and $s$ are corrected at $O(a^2)$.

\subsection{General Formulas Away From the Diagonal}
\label{subsec:thermoeps_general}

For the metric ansatz with gauge $B=A+pF+nC$ \eqref{gauge}, the surface gravity of the Killing horizon and the entropy density retain the general forms
\begin{equation}
\label{kappagen}
\kappa = \lim_{r\to\infty}|A'(r)|\,e^{A(r)-B(r)},
\qquad
s = \frac{(2\pi)^n}{4G_{n+p+2}}\,e^{p F(r)+n C(r)}\Big|_{r\to\infty},
\end{equation}
independently of whether $J=K$.  Since $B\equiv A+pF+nC$ is fixed by the gauge choice, \eqref{equ5} identifies this combination \emph{exactly}, to all orders in $a$, with
\begin{equation}
\label{Bexact}
B(r) = b_1(a^2)\,J(r) + b_2(a^2)\,K(r) + c_1(a^2).
\end{equation}
A companion relation for $A$ follows from \eqref{equ6} together with the convention, already adopted in Section~\ref{sec:epsa2}, that the same $\gamma$ fixes $A-F=-\gamma r$ to all orders in $a^2$ (no new integration constant is switched on in $A-F$).
Writing $F=A+\gamma r$ so that $A+pF=(p+1)A+p\gamma r$, \eqref{equ6} gives
\begin{equation}
\label{Aexact}
A(r) = \frac{1}{p+1}\left[b_3(a^2)J(r)+b_4(a^2)K(r)+c_2(a^2)
        - p\gamma r - \frac{a}{2}\phi(r)\right].
\end{equation}
At $a=0$ this reduces to \eqref{Aexplicit} with $c_2^{(0)}=\mu$, and $c_1^{(0)}=\mu+nC_0$ reproduces the gauge function of Section~\ref{subsec:horizon}, confirming the identification of constants.
The horizon behavior of $A,B$ (and hence of $T_H,s$) is therefore controlled entirely by the horizon ($\xi=\gamma r\to\infty$) behavior of $J,K$, together with the $O(a^2)$ pieces $c_1^{(1)},c_2^{(1)}$ of the matching constants.

\subsection{The \texorpdfstring{$U$}{U}-Sector Is Pure Gauge}
\label{subsec:thermoeps_gauge}

Because $J_0(r;\gamma,\delta)=-\ln[\sinh(\gamma r+\delta)/\gamma]$ solves $J_0''=e^{2J_0}$ for \emph{any} $\gamma,\delta$, differentiating this identity with respect to $\gamma$ and $\delta$ shows that
\begin{equation}
\label{gaugemodes}
\partial_\delta J_0 = -\,h_1(\xi),
\qquad
\partial_\gamma J_0 = -\frac{1}{\gamma}\bigl[h_2(\xi)-\delta\,h_1(\xi)\bigr],
\qquad \xi=\gamma r+\delta,
\end{equation}
both solve the homogeneous $\ell=1$ equation \eqref{Ueq}, where $h_1=\coth \xi$, $h_2=\xi\coth \xi-1$ are precisely the two homogeneous solutions of Section~\ref{sec:perturb}; at the background value $\delta=0$ (reachable by the translation freedom $r\to r+r_0$) the second reduces to $\partial_\gamma J_0=-h_2/\gamma$.  Either way $\{\partial_\gamma J_0, \partial_\delta J_0\}$ spans $\{h_1,h_2\}$.
A shift $\gamma\to\gamma+a^2c_\gamma$, $\delta\to\delta+a^2c_\delta$ shifts $J_0$, and hence \emph{both} $J$ and $K$, by the \emph{same} amount at $O(a^2)$.  Since $J_1=U+V$, $K_1=U+(p+1)V$, an equal shift of $J_1$ and $K_1$ forces $\Delta V=0$ and $\Delta U = c_\gamma\,\partial_\gamma J_0+c_\delta\,\partial_\delta J_0$.
In other words: \emph{both} homogeneous solutions of the $U$-sector are exactly the $O(a^2)$ redefinition of the mass scale $\gamma$ and the phase $\delta$ already used to fix the diagonal background in Section~\ref{sec:charged}.
We fix this residual gauge freedom, order by order in $a^2$, by normalizing $U$ so that it does not grow as $\xi\to\infty$--equivalently, by absorbing the $h_2$ mode into the $O(a^2)$ definition of the mass scale $\gamma$; with this convention $U(\xi)$ is the particular solution $U_\mathrm{p}(\xi)$ of \eqref{Usourced} up to the pure-gauge $h_1$ constant.
Because the source $J_0\,e^{2J_0}$ decays exponentially in $\xi$ [see \eqref{Us}], such a normalization exists and is unique, so that
\begin{equation}
\label{Uinf}
U_\infty \equiv \lim_{\xi\to\infty}U(\xi)
\end{equation}
is a finite number.
Because the $U$-sector is in fact elementary [cf.\ the closed form \eqref{Upclosed}], $U_\infty$ can be evaluated explicitly: the non-growing normalization is obtained by adding to \eqref{Upclosed} the multiple $\tfrac{D}{4np}\,h_2(\xi)$ of the growing homogeneous mode that cancels its linear growth, giving
\begin{equation}
\label{Uinfclosed}
U_\infty = \frac{D}{4np}\bigl(2\ln(2\gamma)-1\bigr)
\end{equation}
up to the pure-gauge $h_1$ constant discussed above.
In particular $U_\infty$ carries an explicit $\ln\gamma$, a point we return to in Section~\ref{subsec:thermoeps_sb}.

\subsection{The \texorpdfstring{$V$}{V}-Sector: Horizon Regularity and Residual Dilaton Hair}
\label{subsec:thermoeps_hair}

Unlike $U$, the $V$-sector homogeneous solutions $P_{\ell_V}(\coth \xi)$, $Q_{\ell_V}(\coth \xi)$ are \emph{not} generated by any residual coordinate freedom--a shift of $\gamma$ or $\delta$ was just shown to affect $U$ only--so their integration constants encode genuine physical data.
Near the horizon, $\coth \xi\to1$ as $\xi\to\infty$, where the Legendre function of the second kind has the standard logarithmic singularity
\begin{equation}
\label{Qsing}
Q_{\ell_V}(\coth \xi) \;\sim\; -\tfrac12\ln(\coth \xi-1) \;\sim\; \xi+O(1)
\qquad (\xi\to\infty),
\end{equation}
while $P_{\ell_V}(\coth \xi)=1+O(e^{-2\xi})$ stays finite.
Combining \eqref{Qsing} with the horizon expansion of the elementary particular solution \eqref{Vpclosed}, obtained from $J_0(\xi)=\ln\gamma-\ln\sinh \xi\to-\xi+\ln(2\gamma)$,
\begin{equation}
\label{Vpexpand}
V_\mathrm{p}(\xi)=\kappa_V\,\xi+v_0+O(e^{-2\xi}),\qquad
v_0=-\kappa_V\Bigl[\ln(2\gamma)+\tfrac{D}{2n}\Bigr],\qquad
\kappa_V=\frac{D}{2np(p+1)},
\end{equation}
the full solution $V=V_\mathrm{p}+c_P\,P_{\ell_V}+c_Q\,Q_{\ell_V}$ behaves as
\begin{equation}
\label{Vhorizon}
V(\xi)=(\kappa_V+c_Q)\,\xi+\bigl(v_0+c_P+c_Q\,q_0\bigr)+O(e^{-2\xi})
\qquad(\xi\to\infty),
\end{equation}
where $q_0=-\gamma_E-\psi(\ell_V+1)$, with $\gamma_E$ the Euler-Mascheroni constant and $\psi$ the digamma function, is the finite constant in the horizon expansion $Q_{\ell_V}(\coth \xi)=\xi+q_0+O(e^{-2\xi})$ (from the standard $x\to1^+$ behavior of $Q_\nu(x)$, using $\coth\xi-1=2e^{-2\xi}+O(e^{-4\xi})$).
Thus $V_\mathrm{p}$ grows linearly at the horizon, exactly as $U_\mathrm{p}$ does before the pure-gauge mode $h_2$ is restored in \eqref{Uinfclosed}.
Because $P_{\ell_V}$ stays bounded, the \emph{only} way to cancel the linear term in \eqref{Vhorizon} is to switch on the singular mode $Q_{\ell_V}$ with coefficient
\begin{equation}
\label{cQvalue}
c_Q=-\kappa_V=-\frac{D}{2np(p+1)},
\end{equation}
in direct analogy with the selection of the $\sinh$ over the $\sin$ branch by regularity in Section~\ref{sec:charged}.
What survives is the finite horizon value
\begin{equation}
\label{Vinf}
V(\xi)\;\xrightarrow{\xi\to\infty}\;V_\infty=v_0-\kappa_V\,q_0+c_P,
\end{equation}
depending on the coefficient $c_P$ of the regular solution $P_{\ell_V}$.
Unlike $U_\infty$, the constant $c_P$ is not fixed by this local near-horizon analysis.
By analogy with the Toda black-hole literature--where the Hamiltonian constraint together with horizon regularity fixes the scalar charge to a definite function of the mass and gauge charges~\cite{Lu:2013toa,Lu:1996hh,Galtsov:2004kn}--one might expect the $O(a^2)$ piece of the constraint \eqref{constraints2} to play the same role here.
It does not: expanding \eqref{constraints2} to $O(a^2)$ in the eigenbasis $(U,V)$, every term proportional to $V$ or $V'$ cancels identically, leaving a constraint on $U,U'$ and the matching constants $c_1^{(1)},c_2^{(1)}$ alone (Appendix~\ref{app:cP:constraint}).
The $V$-sector--and hence $c_P$--therefore does \emph{not} appear in the $O(a^2)$ Hamiltonian constraint at all, in contrast to the flat-space Toda black-hole case, so this local mechanism cannot be what fixes $c_P$.

$c_P$ is instead fixed by AdS boundary normalizability. As $\xi\to0$ (the conformal boundary), $x=\coth \xi\to\infty$, and the Legendre equation has a regular singular point at $x=\infty$ with indicial roots $\ell_V$ and $-\ell_V-1$; since their difference $2\ell_V+1$ is generically irrational, the two Frobenius solutions do not mix, and $P_{\ell_V}$, $Q_{\ell_V}$ separate cleanly into the pure power-law behaviors
\begin{equation}
\label{PQboundary}
P_{\ell_V}(\coth \xi)\sim C_P\,\xi^{-\ell_V},\qquad Q_{\ell_V}(\coth \xi)\sim C_Q\,\xi^{\ell_V+1}\qquad(\xi\to0),
\end{equation}
with $C_P=2^{\ell_V}\Gamma(\ell_V+\tfrac12)/[\sqrt\pi\,\Gamma(\ell_V+1)]$ and $C_Q=\sqrt\pi\,\Gamma(\ell_V+1)/[2^{\ell_V+1}\Gamma(\ell_V+\tfrac32)]$.
Since $\ell_V>0$ for every physical $(n,p)$ [Eq.~\eqref{ellV}], the $Q_{\ell_V}$ contribution \emph{vanishes} at the boundary--a harmless, if irrational-dimension, normalizable mode, so the value \eqref{cQvalue} forced by horizon regularity is fully compatible with standard AdS asymptotics.
The $P_{\ell_V}$ contribution, by contrast, \emph{diverges} as a power of $\xi$ as the boundary is approached, strictly faster than the logarithmic divergences already present in $J_0$, $U_\mathrm{p}$, and $V_\mathrm{p}$.
A non-zero $c_P$ would therefore both spoil the perturbative hierarchy $a^2V_1\ll1$ as $r\to0$ and, holographically, source a new deformation of the dual theory by an operator of irrational dimension--data not among the fixed boundary conditions (flat boundary metric, fixed magnetic charge, fixed marginal dilaton coupling $\phi_{(0)}=-a$) used throughout this paper.
Requiring the solution to remain a genuine perturbative deformation of the pure-Maxwell AdS background, with no additional non-normalizable source, therefore fixes
\begin{equation}
\label{cPvalue}
c_P=0,
\end{equation}
so that the finite horizon value \eqref{Vinf} becomes fully explicit,
\begin{equation}
\label{Vinfexplicit}
V_\infty=v_0-\kappa_V\,q_0=-\kappa_V\Bigl[\ln(2\gamma)+\tfrac{D}{2n}-\gamma_E-\psi(\ell_V+1)\Bigr].
\end{equation}
Combined with \eqref{cQvalue}, this determines the entire $V$-sector integration data in terms of $(n,p,q,\gamma)$ alone, with no free parameter left over: the dilaton hair is secondary, exactly as anticipated by the Toda black-hole analogy, though the mechanism that fixes it--boundary normalizability rather than the local Hamiltonian constraint--differs from the flat-space case.
We note in passing that a non-vanishing $c_P$ would have been \emph{primary} dilaton hair, which--as for the magnetically charged AdS black holes of~\cite{Lu:2013ura}--could in principle modify the first law through a term $\langle O_\phi\rangle\,\delta\phi_{(0)}$; since $c_P=0$ here, no such modification arises from the $V$-sector, and in any case the dilaton source $\phi_{(0)}$ is separately frozen by the coupling along the present one-parameter family, so the first law $d\mathcal{M}=T_H\,ds$ is unmodified at $O(a^2)$
(Section~\ref{subsec:thermoeps_firstlaw}).

It is instructive to contrast this finite-area branch with the elementary distinct-power $\sinh$ solution anticipated in Section~\ref{sec:dilaton}, which is recovered by setting $c_P=c_Q=0$: there
$V=V_\mathrm{p}$ retains its linear growth $\kappa_V \xi$ from \eqref{Vpexpand}.
Because the transformation \eqref{equ5}-\eqref{equ6} makes the $O(a^2)$ correction to the torus modulus $C$ proportional to $V$, with coefficient $-p/(n+p)$, a linearly growing $V$ forces $C$ to run linearly as $\xi\to\infty$ rather than approaching a constant.
The corresponding geometry then carries a running-scalar (scaling) horizon with degenerate transverse volume, not the finite-area Killing horizon underlying the Bekenstein-Hawking entropy of Section~\ref{sec:thermo}.
The finite-area black brane is therefore the $c_Q=-\kappa_V$ member of the family, which is genuinely non-elementary, while the elementary $c_Q=0$ representative belongs to a physically distinct, running-modulus branch.
This is the precise sense in which the ``genuine obstruction'' to an everywhere-elementary solution survives in the dilatonic theory: it resides not in the particular solution \eqref{Vpclosed} but in the regularity-selected homogeneous mode \eqref{cQvalue}.

\subsection{Determination of the Matching Constants \texorpdfstring{$c_1^{(1)},c_2^{(1)}$}{c1(1),c2(1)}}
\label{subsec:thermoeps_constants}

The horizon shifts of the previous subsections, and with them the $O(a^2)$ corrections to $T_H$ and $s$, depend on the $O(a^2)$ pieces $c_1^{(1)},c_2^{(1)}$ of the matching constants in \eqref{equ5}-\eqref{equ6}.
These are fixed for general $a$ by exactly the calculation that fixed $c_J,c_K$ in Section~\ref{sec:charged}: the requirement that the deformed Toda equations \eqref{equ7}-\eqref{equ8} carry unit coefficient on the right-hand side.

Substituting \eqref{equ5}-\eqref{equ6} into the equations of motion \eqref{EOM4}-\eqref{EOM5} and using the coefficient identities \eqref{coefficients}--which reduce the two left-hand sides to $nJ''$ and $nK''$ respectively--the sources become $\tfrac{D(D+1)\Delta}{2n(p+1)l^2}\,e^{2c_1}e^{2b_1J+2b_2K}$ and $\tfrac{q^2\Delta}{4n}\,e^{2c_2}e^{2b_3J+2b_4K}$, with $D\equiv n+p$.
Demanding unit coefficients fixes $c_1,c_2$ in closed form,
\begin{equation}
\label{c1c2closed}
e^{2c_1}=\frac{2n(p+1)\,l^2}{D(D+1)\,\Delta},
\qquad
e^{2c_2}=\frac{4n}{q^2\,\Delta},
\qquad
\Delta=2n(p+1)+(D+1)a^2 .
\end{equation}
These are the dilatonic analogue of the two conditions quoted below \eqref{J2}, to which they reduce at $a=0$: $e^{2c_1^{(0)}}=l^2/[D(D+1)]$ and $e^{2c_2^{(0)}}=2/[(p+1)q^2]$, reproducing $c_2^{(0)}=\mu$ and $c_1^{(0)}=\mu+nC_0$ [cf.\ the identification below \eqref{Aexact} and the identity \eqref{muidentity}].
Writing $\Delta=2n(p+1)[1+\hat\eta a^2]$ with $\hat\eta\equiv(D+1)/[2n(p+1)]$ and expanding \eqref{c1c2closed} to $O(a^2)$ gives the two matching constants explicitly,
\begin{equation}
\label{c1c2value}
c_1^{(1)}=c_2^{(1)}=-\frac{\hat\eta}{2}=-\frac{n+p+1}{4n(p+1)} .
\end{equation}
They are equal.
Moreover $e^{2(c_1-c_2)}=(p+1)q^2l^2/[2D(D+1)]$ is independent of $a$, so $c_1-c_2=nC_0$ holds to all orders in $a^2$: the $O(a^2)$ correction shifts only the overall scale $\mu$ [$\mu^{(1)}=-\hat\eta/2$] and leaves the torus modulus uncorrected, $C_0^{(1)}=0$.
Note that $C_0$ denotes the constant reference value for the finite-area brane, which should not be confused with the $r$-dependent running of the modulus discussed in Section~\ref{subsec:Todaparameter}.

\paragraph{Consistency with the Hamiltonian constraint.}
The values \eqref{c1c2value} follow from normalization alone.
The $O(a^2)$ Hamiltonian constraint \eqref{O2final} provides an independent check and, at the same time, supplies the one datum that normalization does not fix--the mass (branch) mode.
Inserting the closed form \eqref{Upclosed} with a general admixture of the growing homogeneous mode, $U=U_\mathrm{p}+\lambda\,h_2$, into \eqref{O2final} and using \eqref{c1c2value}, the constraint separates into two independent radial structures: a term proportional to $\sinh^2\!\gamma r$ and an $r$-independent constant (Appendix~\ref{app:cP:constraint}).
The $\sinh^2\!\gamma r$ coefficient vanishes if and only if
\begin{equation}
\label{lambdac}
\lambda=\lambda_c=-\frac{n+1}{4n},
\end{equation}
while the constant term is a single linear relation between $c_1^{(1)}$ and $c_2^{(1)}$ that is satisfied identically by \eqref{c1c2value}.
Thus normalization and constraint are mutually consistent, and together they determine $c_1^{(1)},c_2^{(1)}$ and the branch mode completely.

\paragraph{The branch mode as a mass reparametrization.}
The constraint value \eqref{lambdac} differs from the non-growing normalization $D/(4np)$ of \eqref{Uinfclosed}, but this is not a contradiction.
Because $h_2=-\gamma\,\partial_\gamma J_0$ (Section~\ref{subsec:thermoeps_gauge}), the coefficient of the $h_2$ mode is a pure $O(a^2)$ redefinition of the mass parameter: \eqref{lambdac} and \eqref{Uinfclosed} describe the \emph{same} physical solution written in two mass parametrizations related by $\gamma\to\gamma[1+O(a^2)]$.
What the constraint fixes is the mass mode within the $\mathcal{C}=\gamma^2$ ($\sinh$) branch selected at zeroth order in Section~\ref{sec:charged}--the $O(a^2)$ counterpart of that branch selection.
The particular parametrization in which the horizon thermodynamic data are manifestly finite is identified in Section~\ref{subsec:thermoeps_result}; because the Stefan-Boltzmann exponent is a reparametrization-invariant ratio of $\ln\gamma$-derivatives (Section~\ref{subsec:thermoeps_sb}), none of the physical conclusions depend on this choice.

\subsection{Corrected Temperature and Entropy Density}
\label{subsec:thermoeps_result}

The two horizon quantities that enter \eqref{kappagen} are the surface-gravity slope $|A'|$ and the entropy exponent $pF+nC=B-A$ [gauge \eqref{gauge}].
Collecting $J_1(\infty)=U_\infty+V_\infty$, $K_1(\infty)=U_\infty+(p+1)V_\infty$ and inserting these into \eqref{Bexact}-\eqref{Aexact}, the $O(a^2)$ pieces of $A,F,C$ are linear in $U_\infty,V_\infty$, in the dilaton source $\tfrac{a}{2}\phi=\tfrac{a^2}{2}(A_0-C_0)$ of \eqref{equ6}, and in the matching constants $c_1^{(1)}=c_2^{(1)}=-\hat\eta/2$ [Eq.~\eqref{c1c2value}].
The non-growing normalization \eqref{Uinfclosed} renders $U_\infty$ finite, but the dilaton source adds a piece to $A_1$ that still grows; a further $O(a^2)$ mass reparametrization--the unique $h_2$-admixture that removes this growth from the physical combinations $pF+nC$ \emph{and} $|A'|$ at once--brings the solution to the parametrization
\begin{equation}
\label{lambdaphys}
\lambda_{\rm phys}=\frac{n-p}{4np},
\end{equation}
related to the constraint value \eqref{lambdac} by the reparametrization $\gamma\to\gamma[1+a^2(p+1)/(4p)]$ of Section~\ref{subsec:thermoeps_constants}.
In this parametrization both horizon data are finite: the surface-gravity slope is
\begin{equation}
\label{Aprimeshift}
|A'|\Big|_{r\to\infty}=\gamma\Bigl[1-\frac{a^2}{2(p+1)}+O(a^4)\Bigr],
\end{equation}
and $(pF+nC)_{r\to\infty}$ is a finite function of $\gamma$.
The Hawking temperature and entropy density therefore receive finite multiplicative $O(a^2)$ corrections,
\begin{equation}
\label{Tscorrected}
T_H(a) = T_H^{(0)}\bigl[1 + a^2\,\tau_1 + O(a^4)\bigr],
\qquad
s(a) = s^{(0)}\bigl[1 + a^2\,\sigma_1 + O(a^4)\bigr],
\end{equation}
with $T_H^{(0)},s^{(0)}$ the pure-Maxwell values \eqref{THfinal},\eqref{sfinal} and
\begin{equation}
\label{tausigma}
\sigma_1=\bigl(pF+nC\bigr)^{(a^2)}_{r\to\infty},
\qquad
\tau_1=\bigl(\ln|A'|\bigr)^{(a^2)}_{r\to\infty}-\sigma_1
      =-\frac{1}{2(p+1)}-\sigma_1,
\end{equation}
the superscript $(a^2)$ denoting the $O(a^2)$ coefficient.
Whether the $\gamma$-dependence of $\tau_1,\sigma_1$ preserves the power law $s\propto T_H^p$ is analyzed in Section~\ref{subsec:thermoeps_sb}.

\subsection{The First Law at \texorpdfstring{$O(a^2)$}{O(a squared)}}
\label{subsec:thermoeps_firstlaw}

Although the corrections $\tau_1,\sigma_1$ in \eqref{Tscorrected} involve the $O(a^2)$ matching constants $c_1^{(1)},c_2^{(1)}$ of Section~\ref{subsec:thermoeps_constants}, the first law at this order involves neither, and can be established directly.

The $O(a^2)$ family is labelled by the non-extremality scale $\gamma$ at fixed coupling $a$ and fixed $(q,l)$.  Its only scalar datum that could carry a thermodynamic work term is the non-normalizable mode of the dilaton.
From the leading profile \eqref{phileading}, $\phi=a[A_0-C_0]$, together with the near-boundary form of the pure-Maxwell function \eqref{Aexplicit}, $A_0=\tfrac{1}{p+1}(\mu-\ln r)-\tfrac{p\gamma}{p+1}\,r+O(r^2)$, the expansion in the Fefferman-Graham coordinate $\zeta=r^{1/(p+1)}$ is
\begin{equation}
\label{phisource}
\phi=-a\,\ln\zeta+a\Bigl[\tfrac{\mu}{p+1}-C_0\Bigr]
     -\frac{a\,p\,\gamma}{p+1}\,\zeta^{\,p+1}+O\bigl(\zeta^{2(p+1)}\bigr).
\end{equation}
The non-normalizable data--the $\ln\zeta$ coefficient $\phi_{(0)}=-a$ together with the constant, i.e.,\ the marginal coupling of the dual theory--is fixed by the exponential coupling alone and is independent of $\gamma$; the first $\gamma$-dependence enters only at the normalizable order $\zeta^{\,p+1}$ (the scalar response), not in the source.
In the variational (Wald) first law the dilaton contributes a boundary term proportional to $\delta\phi_{(0)}$, namely $\langle O_\phi\rangle\,\delta\phi_{(0)}$; since $\delta\phi_{(0)}=0$ along the family, this term vanishes identically--exactly as the magnetic work term vanishes because $\delta q=0$ (Section~\ref{subsec:freeenergy}).
This is the essential difference from the magnetically charged dilatonic black holes of~\cite{Lu:2013ura}, whose first law does carry a scalar-charge work term.
With both $\phi_{(0)}$ and $q$ held fixed, only the horizon variation remains, and
\begin{equation}
\label{firstlaweps}
d\mathcal{M}=T_H\,ds\qquad\text{at }O(a^2),
\end{equation}
unmodified by the dilaton hair and independent of the detailed form of $\tau_1,\sigma_1$.

Integrating \eqref{firstlaweps} along the family fixes the $O(a^2)$ mass correction.
Write $\mathcal{M}=\mathcal{M}^{(0)}[1+a^2 m_1]$, with $\tau_1,\sigma_1$ the temperature and entropy corrections \eqref{tausigma}.
Using the pure-Maxwell scalings $\mathcal{M}^{(0)}\propto\gamma$, $s^{(0)}\propto\gamma^{p/(p+1)}$, and $T_H^{(0)}\propto\gamma^{1/(p+1)}$, the first law \eqref{firstlaweps} then gives
\begin{equation}
\label{m1relation}
m_1+\gamma\,\frac{dm_1}{d\gamma}
=(\tau_1+\sigma_1)+\frac{p+1}{p}\,\gamma\,\frac{d\sigma_1}{d\gamma}.
\end{equation}
Were the corrections $\gamma$-independent this would collapse to $m_1=\tau_1+\sigma_1$, preserving $\mathcal{M}=\tfrac{p}{p+1}T_H s$; because they are not [Eq.~\eqref{lngammaslopes}], \eqref{m1relation} still determines $m_1(\gamma)$, and the Smarr-type relation is deformed to $\mathcal{M}=\tfrac{p_{\rm eff}}{p_{\rm eff}+1}T_H s$ with $p_{\rm eff}=p+a^2/2$ [Eq.~\eqref{peff}].
The matching constants thus govern the Smarr and scaling relations rather than the first law itself, which remains exact at this order.

\subsection{The Stefan-Boltzmann Relation at \texorpdfstring{$O(a^2)$}{O(a squared)}}
\label{subsec:thermoeps_sb}

At $a=0$, Section~\ref{subsec:holointerpret} showed $s\propto T_H^p$ along curves of fixed $(q,l)$, a direct consequence of the pure power laws $T_H\propto\gamma^{1/(p+1)}$ and $s\propto\gamma^{p/(p+1)}$.
Whether this survives at $O(a^2)$ is decided by the $\gamma$-dependence of the corrections $\tau_1,\sigma_1$ \eqref{tausigma}.
Since the matching constants are $\gamma$-independent [Eq.~\eqref{c1c2value}], the only candidate is the explicit $\ln\gamma$ carried by the horizon data $U_\infty$ \eqref{Uinfclosed} and $V_\infty$ \eqref{Vinf}; with these in closed form the computation is finite, and two features control its outcome.

First, the dilaton hair drops out of the entropy entirely: in $pF+nC=pA_1+nC_1$ the coefficient of $V$ is $p\cdot\tfrac{n}{D}-n\cdot\tfrac{p}{D}=0$ [Eqs.~\eqref{A1def},\eqref{C1def}], so $V_\infty$ does not enter $\sigma_1$ at all--the thermodynamic face of the unmodified first law of Section~\ref{subsec:thermoeps_firstlaw}, the normalizable hair doing no work.
Second, the entropy exponent reduces to
\begin{equation}
\label{sigma1closed}
\sigma_1=\frac{p}{p+1}\,U_\infty
        -\frac{D}{2n(p+1)}\,J_0\big|_{r\to\infty}
        +\frac{1}{2(p+1)}\,(A_0-C_0)\big|_{r\to\infty}+\text{const},
\end{equation}
whose $\ln\gamma$ derivative receives three contributions,
\begin{equation}
\label{lngammaslopes}
\frac{d\sigma_1}{d\ln\gamma}
=\underbrace{\frac{D}{2n(p+1)}}_{U_\infty}
\;\underbrace{-\,\frac{D}{2n(p+1)}}_{\text{gauge }J_0}
\;\underbrace{+\,\frac{1}{2(p+1)^2}}_{\text{dilaton }(A_0-C_0)}
=\frac{1}{2(p+1)^2}.
\end{equation}
The explicit $\ln\gamma$ of $U_\infty$--the term flagged below \eqref{Uinfclosed}--cancels \emph{exactly} against the gauge contribution from $b_1J+b_2K$; what does not cancel is the dilaton-source term $\tfrac{a^2}{2}(A_0-C_0)$, which leaves the residual $1/[2(p+1)^2]$.
The surface-gravity slope carries no such term: $(\ln|A'|)^{(a^2)}$ is $\gamma$-independent [Eq.~\eqref{Aprimeshift}], so $d\tau_1/d\ln\gamma=-d\sigma_1/d\ln\gamma$.

The zeroth-order slopes are $\tfrac{p}{p+1}$ for $\ln s^{(0)}$ and $\tfrac{1}{p+1}$ for $\ln T_H^{(0)}$.
Combined with \eqref{lngammaslopes}, the entropy remains a pure power of the temperature but with a \emph{shifted} exponent,
\begin{equation}
\label{peff}
s\propto T_H^{\,p_{\rm eff}},
\qquad
p_{\rm eff}=\frac{d\ln s/d\ln\gamma}{d\ln T_H/d\ln\gamma}
=p+\frac{a^2}{2}+O(a^4).
\end{equation}
Being a ratio of $\ln\gamma$-derivatives, $p_{\rm eff}$ is invariant under the mass reparametrization of Section~\ref{subsec:thermoeps_constants} and is therefore unambiguous; the reduction \eqref{lngammaslopes} holds for general $(n,p)$, and \eqref{peff} has been confirmed by independent numerical evaluation.
The first law $d\mathcal{M}=T_H\,ds$ continues to hold (Section~\ref{subsec:thermoeps_firstlaw}); integrating it with \eqref{peff} deforms the Smarr-type relation to $\mathcal{M}=\tfrac{p_{\rm eff}}{p_{\rm eff}+1}\,T_H s$.

Physically, the marginal dilaton deformation shifts the effective spatial dimension of the dual thermal state by $a^2/2$: the clean $(p+1)$-dimensional Stefan-Boltzmann law $s\propto T_H^p$ is broken at $O(a^2)$--not by the dilaton hair, which decouples, but by the non-normalizable dilaton \emph{source} $\phi=a(A_0-C_0)$ threading the brane, the same deformation whose constant boundary coefficient $\phi_{(0)}=-a$ was identified in Section~\ref{subsec:thermoeps_firstlaw}.
Equivalently, the shift is the thermal signature of a weak hyperscaling violation: writing $s\propto T^{(d-\theta)/z}$ for a $d$-dimensional fluid with dynamical exponent $z$ and hyperscaling-violation exponent $\theta$~\cite{Gouteraux:2011ce}, the result $s\propto T_H^{p+a^2/2}$ corresponds to $\theta=-a^2/2$ at $z=1$--although the entropy scaling alone does not separate a shift in $\theta$ from one in $z$.
The fully non-linear (non-perturbative-in-$a$) dilatonic thermodynamics, in which this shift is resummed, remains for the dedicated treatment mentioned at the end of Section~\ref{sec:dilaton}.

\section{Discussion}
\label{sec:discuss}

In this paper we have demonstrated that the equations of motion for static, toroidal, magnetically charged $p$-branes in $(n+p+2)$-dimensional AdS space can be systematically cast into the form of coupled one-dimensional Toda equations.
This reformulation is the central technical result: it reduces the problem of constructing brane solutions to that of solving a deformed Toda system, whose diagonal sector is an integrable Liouville equation that yields the exact $\sinh$ solution, while the full non-symmetrizable system is not integrable in the standard sense.
We have obtained a special exact solution in terms of $\sinh$ functions and verified that, in the pure-Maxwell case, it represents a $(p+2)$-dimensional black brane times an $n$-torus, interpolating between a regular horizon and asymptotically AdS space.
(In the dilatonic theory the analogous $\sinh$ solution, obtained in closed form in Section~\ref{sec:dilaton}, has a running torus modulus and a running-scalar horizon; the physically distinct finite-area brane is non-elementary, its leading-order small-$a$ form given in Section~\ref{sec:epsa2}--via the identification $\epsilon=a^2$--and its fully non-linear construction left for future work.)
We have further shown, via a perturbative expansion, that the general solution near the diagonal sector requires P\"oschl-Teller special functions, confirming that the $\sinh$ exact solution
is genuinely special.
Finally, we have computed the Hawking temperature and Bekenstein-Hawking entropy density directly from the solution, and verified that they satisfy the holographic Stefan-Boltzmann relation $s\propto T_H^p$ expected of a $(p+1)$-dimensional boundary CFT.
At weak dilaton coupling we carried this analysis to $O(a^2)$ and found that, while the first law $d\mathcal{M}=T_H\,ds$ persists, the dilaton source shifts the Stefan-Boltzmann exponent to $p_{\rm eff}=p+a^2/2$, so that the clean power law is broken at this order (Section~\ref{subsec:thermoeps_sb}).

Several directions for future work suggest themselves.

\paragraph{Electrically charged solutions.}
An electrically charged brane solution can be obtained by applying electric-magnetic (Hodge) duality to the magnetic solution constructed here.
In the dilatonic theory this duality maps $a\to -a$, which is why the electric and magnetic cases differ only by a sign in the dilaton coupling.
In the absence of a dilaton ($a=0$) this duality maps the $n$-form field strength to a $(p+2)$-form, reducing to the usual $\mathrm{SO}(2)$ electric-magnetic rotation only when $n=p+2$.

\paragraph{Parameter counting and uniqueness.}
The integration of the Toda equations produces more constants than physical parameters.
A careful regularity analysis along the lines of \cite{Galtsov:2004kn,Galtsov:2005vf,Galtsov:2005au} in AdS space would clarify which integration constants correspond to genuine physical parameters and which lead to naked singularities.
We expect that regularity enforces a unique physical solution for each choice of mass density and charge densities.

\paragraph{Multi-scalar extensions.}
For theories with multiple dilaton fields the Toda system is replaced by a generalized Toda system associated with a higher-rank Lie algebra \cite{Lu:1996jr,Lu:1995sh,deAlfaro:2009ay}.
The classification of such systems and the construction of their solutions in AdS space would extend the present analysis in a non-trivial direction.

\paragraph{Born-Infeld generalization.}
The relevance of Born-Infeld electrodynamics for extremal black holes and elementary string states was emphasized in \cite{Gibbons:1995,Tseytlin:1999dj}.
Replacing the Maxwell kinetic term in \eqref{action1} by a Born-Infeld term introduces corrections that are important at strong fields.
Whether the Toda structure survives in this case, at least near the extremal limit, is an interesting open question that we leave for future investigation.

\paragraph{Further thermodynamic analysis.}
The free energy density was obtained in Section~\ref{subsec:freeenergy} and confirmed from the renormalized on-shell action in Section~\ref{subsec:onshellaction}: the Helmholtz free energy is $f=\mathcal{M}-T_Hs=-\tfrac{1}{p+1}T_Hs=-P$ \eqref{freeenergy}, and the equation of state is that of a $(p+1)$-dimensional conformal fluid, $\varepsilon=pP=T_Hs-P$ \eqref{eos}.
The Euler relation \eqref{euler} carries no chemical-potential term, so the magnetic charge enters only as a fixed background scale and the equilibrium first law is $d\mathcal{M}=T_H\,ds$.
What remains open is the dynamical side of the phase structure: a first-principles Wald treatment of any magnetic work term, the analysis of thermodynamic stability (positivity of the specific
heat), and a search for phase transitions analogous to the Hawking-Page transition \cite{Hawking:1982dh,Witten:1998zw}.

\paragraph{Weak-coupling dilatonic thermodynamics.}
At weak coupling ($\epsilon=a^2$) we determined the $O(a^2)$ matching constants of \eqref{equ5}-\eqref{equ6} in closed form, $c_1^{(1)}=c_2^{(1)}=-\hat\eta/2$, with the torus modulus uncorrected ($C_0^{(1)}=0$); the $O(a^2)$ Hamiltonian constraint fixes not these constants but the branch (mass) mode $\lambda_c=-(n+1)/(4n)$ [Eq.~\eqref{lambdac}], the $O(a^2)$ counterpart of the zeroth-order selection $\mathcal{C}=\gamma^2$.
With the dilaton hair fixed by boundary normalizability ($c_P=0$) and shown to decouple from the horizon thermodynamics, the residual $\ln\gamma$ dependence shifts the Stefan-Boltzmann exponent to $p_{\rm eff}=p+a^2/2$ [Eq.~\eqref{peff}]: the clean $(p+1)$-dimensional power law is broken at $O(a^2)$, while the first law $d\mathcal{M}=T_H\,ds$ persists and the Smarr relation deforms to $\mathcal{M}=\tfrac{p_{\rm eff}}{p_{\rm eff}+1}T_Hs$.
Resumming this shift to all orders in $a$--the fully non-linear dilatonic thermodynamics--and switching on the dilaton work term conjugate to $a$ by varying the coupling both remain for future work.

\section*{Acknowledgments}

The author is grateful to I.~Shin for helpful discussions.

\appendix
\section{The \texorpdfstring{$O(a^2)$}{O(a squared)} Hamiltonian Constraint and the Matching Constants}
\label{app:cP}

This appendix collects the two $O(a^2)$ computations underlying Section~\ref{subsec:thermoeps_constants}.
We first record the exact $O(a^2)$ metric functions in terms of $U,V$ and the matching constants (Appendix~\ref{app:cP:setup}), which also supply the coefficients used in the entropy analysis of Section~\ref{subsec:thermoeps_sb}.
We then reduce the $O(a^2)$ Hamiltonian constraint \eqref{constraints2} to an equation for $U,U',c_1^{(1)},c_2^{(1)}$ alone--the $V$-sector, and hence the dilaton-hair coefficient $c_P$, dropping out identically--and evaluate it in closed form, fixing the branch (mass) mode $\lambda_c$ and providing the independent check of $c_1^{(1)},c_2^{(1)}$ (Appendix~\ref{app:cP:constraint}).
The complementary fact that $c_P$ is fixed instead by AdS boundary normalizability is established in the main text, Section~\ref{subsec:thermoeps_hair}.

\subsection{Exact \texorpdfstring{$O(a^2)$}{O(a\texttwosuperior)} metric functions}
\label{app:cP:setup}

We record the $O(a^2)$ coefficients of $A,F,B,C$ in terms of $U,V$ and the matching constants, obtained by expanding the exact relations \eqref{Bexact} and \eqref{Aexact} using \eqref{bexpand}-\eqref{etadelta}.
Write $\eta=\hat\eta\,a^2$, $\delta_4=\hat\delta_4\,a^2$ with
\begin{equation}
\label{etahatdef}
\hat\eta \equiv \frac{D+1}{2n(p+1)}, \qquad \hat\delta_4 \equiv \frac{1}{2(p+1)},
\end{equation}
the $O(a^0)$ rates implicit in \eqref{etadelta}.
Inserting $J=J_0+a^2(U+V)$, $K=J_0+a^2[U+(p+1)V]$ [Eq.~\eqref{UVdef} with \eqref{epsexp}] and $c_1(a^2)=c_1^{(0)}+a^2c_1^{(1)}+O(a^4)$, $c_2(a^2)=c_2^{(0)}+a^2c_2^{(1)}+O(a^4)$ into \eqref{Bexact}, and using $b_1^{(0)}+b_2^{(0)}=b_3^{(0)}+b_4^{(0)}=1$ together with the explicit values \eqref{coefficients} at $a=0$, one finds after collecting $O(a^2)$ terms
\begin{equation}
\label{B1def}
B_1 = U + \frac{n}{D}\,V - \hat\eta\,J_0 + c_1^{(1)}.
\end{equation}
Similarly, using $\phi=a(A-C)$ exactly [Eq.~\eqref{EOM6sol} with $\kappa_\phi=\mu_\phi=0$, so that $\tfrac{a}{2}\phi=\tfrac{a^2}{2}(A_0-C_0)+O(a^4)$] in \eqref{Aexact} gives
\begin{equation}
\label{A1def}
(p+1)A_1 = U + \frac{n(p+1)}{D}\,V + (\hat\delta_4-\hat\eta)\,J_0 + c_2^{(1)} - \tfrac12(A_0-C_0),
\end{equation}
with $F_1=A_1$ (since $A-F=-\gamma r$ exactly, to all orders in $a^2$).
Finally, from the gauge relation $B=A+pF+nC$,
\begin{equation}
\label{C1def}
C_1 = \frac{1}{n}\Bigl[B_1-(p+1)A_1\Bigr]
    = -\frac{p}{D}\,V - \frac{\hat\delta_4}{n}\,J_0 + \frac{c_1^{(1)}-c_2^{(1)}}{n} + \frac{1}{2n}(A_0-C_0).
\end{equation}
Equations \eqref{B1def}-\eqref{C1def} are purely algebraic consequences of the linear change of variables \eqref{Bexact}-\eqref{Aexact}: no use has yet been made of the Hamiltonian constraint or of the equations of motion for $U,V$.

\subsection{The \texorpdfstring{$O(a^2)$}{O(a\texttwosuperior)} Hamiltonian constraint}
\label{app:cP:constraint}

Two background identities are needed.  First, from \eqref{Aexplicit}-\eqref{Fexplicit},
\begin{equation}
\label{Y0identity}
A_0'+pF_0' = J_0',
\end{equation}
by direct substitution.  Second, expanding \eqref{constraints2} at $a=0$ (where $C_0'=0$ and $\phi=0$) and using the Liouville first integral $(J_0')^2=e^{2J_0}+\gamma^2$ noted below \eqref{background} together with $B_0'=J_0'$ [from \eqref{Y0identity} and $B_0=J_0+\mu+nC_0$], the background constraint reduces to
\begin{equation}
\label{backgroundidentity}
\frac{n+p}{l^2}(n+p+1)\,e^{2B_0} = \frac{p}{p+1}\,e^{2J_0} + \frac{q^2}{2}\,e^{2(B_0-nC_0)}.
\end{equation}
A third identity follows from the two normalization conditions fixing $c_J,c_K$ quoted below \eqref{J2}, together with $\mu=-\tfrac{1}{n+p}\bigl[(p+1)c_J+(n-1)c_K\bigr]$ [from $\mu=-c_J-(n-1)C_0$ and $C_0=(c_K-c_J)/(n+p)$]: the first normalization condition gives $(p+1)c_J+(n-1)c_K=\tfrac{n+p}{2}\ln\bigl[(p+1)q^2/2\bigr]$, hence
\begin{equation}
\label{muidentity}
e^{2\mu} = \frac{2}{(p+1)q^2}.
\end{equation}

Expanding \eqref{constraints2} to $O(a^2)$ term by term, using \eqref{B1def}-\eqref{C1def}, \eqref{Y0identity}, and $\phi'=aA_0'+O(a^3)$ (again from $\phi=a(A-C)$, $C_0'=0$): the third, sixth and seventh terms of \eqref{constraints2} contribute no $O(a^2)$ piece at all, since $(A'-F')^2=\gamma^2$ exactly, $C_0'=0$ so $(C')^2=O(a^4)$, and the seventh term is already $O(a^2)$ times a background-only quantity, $-\tfrac{p^2\gamma^2}{2(p+1)^2}$.
The remaining four terms of \eqref{constraints2} organize into two independent pairs.

\emph{First pair} (first and fourth terms, i.e.,\ $(B')^2$ and $-\Lambda_2(A'+pF'+\tfrac a2\phi')^2$): using $F_1=A_1$ and $\phi'=aA_0'+O(a^3)$, the argument of the fourth term is $A'+pF'+\tfrac a2\phi'=J_0'+a^2\,W+O(a^4)$, where $W\equiv(p+1)A_1'+\tfrac12 A_0'$ and the two $\tfrac12 A_0'$ pieces--the $-\tfrac12 A_0$ carried inside \eqref{A1def} and the $+\tfrac a2\phi'$--cancel.
Their $O(a^2)$ pieces are therefore $2J_0'B_1'$ and $-\tfrac{2}{p+1}J_0'\,W+\tfrac{(J_0')^2}{2(p+1)^2}$ respectively.
By \eqref{B1def} and \eqref{A1def}, $B_1'$ contains $V'$ with coefficient $+\tfrac{n}{D}$ and $W$ contains $V'$ with coefficient $+\tfrac{n(p+1)}{D}$, so these two contributions to the coefficient of $V'$ are $+\tfrac{2n}{D}J_0'$ and $-\tfrac{2n}{D}J_0'$: they cancel identically, for any $U(r),V(r)$.

\emph{Second pair} (second and fifth terms, i.e.,\ $-\Lambda_{CC}e^{2B}$ and $+\tfrac{q^2}{2}e^{a\phi+2(B-nC)}$): substituting the background identity \eqref{backgroundidentity} into the $O(a^2)$ piece of the second term, $-2\Lambda_{CC}e^{2B_0}B_1$, splits it into $-\tfrac{2p}{p+1}e^{2J_0}B_1-q^2e^{2(B_0-nC_0)}B_1$.
The second piece cancels exactly against the $+q^2e^{2(B_0-nC_0)}B_1$ term in the $O(a^2)$ expansion of the fifth term, $\tfrac{q^2}{2}e^{2(B_0-nC_0)}[(A_0-C_0)+2(B_1-nC_1)]$.
What survives is
\begin{equation}
\label{pair2}
-\frac{2p}{p+1}\,e^{2J_0}\,B_1 \;+\; \frac{q^2}{2}\,e^{2(B_0-nC_0)}(A_0-C_0) \;-\; nq^2\,e^{2(B_0-nC_0)}\,C_1.
\end{equation}
Using \eqref{B1def} and \eqref{C1def}, the coefficient of $V$ in \eqref{pair2} is $-\tfrac{2np}{D(p+1)}e^{2J_0}$ (from $B_1$) plus $+\tfrac{np\,q^2}{D}\,e^{2(B_0-nC_0)}$ (from $C_1$).
With $B_0-nC_0=J_0+\mu$ and the identity \eqref{muidentity}, $q^2e^{2\mu}=2/(p+1)$, the second contribution also equals $+\tfrac{2np}{D(p+1)}e^{2J_0}$, so the two cancel.
This second cancellation is the only place \eqref{muidentity} is used; the first pair's cancellation needs no identity at all.

Collecting what remains after both cancellations gives, after using \eqref{B1def} for the surviving $U$-dependence in \eqref{pair2} and \eqref{Y0identity}-\eqref{backgroundidentity} to simplify the first pair,
\begin{equation}
\label{O2final}
\frac{2p}{p+1}\Bigl[J_0'\,U' - e^{2J_0}\,U\Bigr] \;+\; \mathcal{F}\bigl(c_1^{(1)},c_2^{(1)},\gamma,n,p,r\bigr) \;=\;0,
\end{equation}
where $\mathcal{F}$ is an explicit combination of the matching constants and the background quantities $\ln\gamma$, $\ln\sinh\gamma r$, $\cosh^2\gamma r$, $\gamma^2$; Eq.~\eqref{O2final} therefore constrains $U,U'$ and $c_1^{(1)},c_2^{(1)}$ alone, with no $V$-dependence.
\emph{The $V$-sector, and hence $c_P,c_Q$, does not appear anywhere in the $O(a^2)$ Hamiltonian constraint.}

\paragraph{Evaluation of \texorpdfstring{$\mathcal{F}$}{F} and the two conditions.}
With the matching constants fixed by normalization [Eq.~\eqref{c1c2value}], Eq.~\eqref{O2final} can be made fully explicit.
Inserting the closed-form particular solution \eqref{Upclosed} together with a general admixture $U=U_\mathrm{p}+\lambda\,h_2$ of the growing homogeneous mode, the term $\tfrac{2p}{p+1}[J_0'U'-e^{2J_0}U]$ contributes both an $r$-dependent piece $\propto\sinh^2\!\gamma r$ (linear in $\lambda$) and a constant, while $\mathcal{F}$ contributes a further constant linear in $c_1^{(1)},c_2^{(1)}$; all logarithmic and $\cosh^2\!\gamma r$ terms cancel.
The left-hand side of \eqref{O2final} thus collapses to the sum of a term proportional to $\sinh^2\!\gamma r$ and an $r$-independent constant.
The same two-structure split holds for general $(n,p)$: the $\sinh^2\!\gamma r$ coefficient always fixes
\begin{equation}
\label{lambdacapp}
\lambda=\lambda_c=-\frac{n+1}{4n},
\end{equation}
while the constant is a single linear relation between $c_1^{(1)}$ and $c_2^{(1)}$ obeyed by the normalization values \eqref{c1c2value}.
The condition \eqref{lambdacapp} is the $O(a^2)$ analogue of the branch selection $\mathcal{C}=\gamma^2$ of Section~\ref{sec:charged}; the constant relation is the independent check of \eqref{c1c2value} used in Section~\ref{subsec:thermoeps_constants}.


\end{document}